\documentclass[11pt]{article}
\usepackage{amsfonts,amsmath,amssymb}
\usepackage{enumerate}
\usepackage{hyperref}
\usepackage{bbm}
\usepackage{nicefrac}
\usepackage[all]{xy}
\usepackage{graphicx}
\usepackage{bm}
\usepackage{makecell}
\usepackage[table]{xcolor}

\usepackage{upgreek}

\usepackage{float}			% NEW Helps position tables right here [H]
\usepackage{lscape}                 % NEW creates pages in landscape format

\usepackage{booktabs}

\newcommand{\be}{\begin{equation}}
\newcommand{\ee}{\end{equation}}

\newcommand{\bea}{\begin{eqnarray}}
\newcommand{\eea}{\end{eqnarray}}

\newcommand{\bes}{\begin{subequations}}
\newcommand{\ees}{\end{subequations}}

\newcommand{\cN}{{\cal N}}

\def\sst#1{{\scriptscriptstyle #1}}

\def\0{{\sst{(0)}}}
\def\1{{\sst{(1)}}}
\def\2{{\sst{(2)}}}
\def\3{{\sst{(3)}}}
\def\4{{\sst{(4)}}}
\def\5{{\sst{(5)}}}
\def\6{{\sst{(6)}}}
\def\7{{\sst{(7)}}}
\def\8{{\sst{(8)}}}

\allowdisplaybreaks  %Introduces page breaks inside \eqnarray%

\usepackage{multirow}
\usepackage{rotating}

\newcommand{\ba}{\begin{align}}
\newcommand{\ea}{\end{align}}

\newcommand{\bse}{\begin{subequations}}
\newcommand{\ese}{\end{subequations}}

\allowdisplaybreaks

\usepackage{lipsum}

\newlength\Colsep
\begin{document}

\makeatletter
\renewcommand{\theequation}{\thesection.\arabic{equation}}
\@addtoreset{equation}{section}
\makeatother

\begin{titlepage}

\begin{flushright}
IFT-UAM/CSIC-20-175 \\
%
%\today
\end{flushright}

\vspace{5pt}

   \begin{center}
   \baselineskip=16pt

   \begin{Large}\textbf{
\hspace{-18pt} Kaluza-Klein fermion mass matrices  \\[8pt]
from Exceptional Field Theory 
and ${\cal N} =1$ spectra
}
   \end{Large}

\vspace{25pt}

{\large  Mattia Ces\`aro$^{1}$ \,and \,  Oscar Varela$^{1,2}$}
		
\vspace{25pt}

	\begin{small}

	{\it $^{1}$ Departamento de F\'\i sica Te\'orica and Instituto de F\'\i sica Te\'orica UAM/CSIC , \\
   Universidad Aut\'onoma de Madrid, Cantoblanco, 28049 Madrid, Spain}  \\

	\vspace{15pt}
	
	{\it $^{2}$ Department of Physics, Utah State University, Logan, UT 84322, USA}     \\	
		
	\end{small}

\vskip 50pt

\end{center}

\begin{center}
\textbf{Abstract}
\end{center}

\begin{quote}

Using Exceptional Field Theory, we determine the infinite-dimensional mass matrices for the gravitino and spin-$1/2$ Kaluza-Klein perturbations above a class of anti-de Sitter solutions of M-theory and massive type IIA string theory with topologically-spherical internal spaces. We then use these mass matrices to compute the spectrum of Kaluza-Klein fermions about some solutions in this class with internal symmetry groups containing SU(3). Combining these results with previously known bosonic sectors of the spectra, we give the complete spectrum about some ${\cal N}=1$ and some non-supersymmetric solutions in this class. The complete spectra are shown to enjoy certain generic features.

\end{quote}

\vfill

\end{titlepage}

\tableofcontents

%%%%%%%%%%%%%%%%%%%%%%%%%%%%%%%%%%%%%%%%%%%%%%%%%%%%%%%%%%%%%%%%%%%%%%%%%%

%%%%%%%%%%%%%%%
%%%%%%%%%%%%%%%

\section{Introduction}

%%%%%%%%%%%%%%%
%%%%%%%%%%%%%%%

By construction, the duality-covariant reformulations \cite{Pacheco:2008ps,Berman:2010is,Hohm:2013pua,Hohm:2013vpa,Hohm:2013uia,Ciceri:2016dmd,Cassani:2016ncu} (see \cite{Berman:2020tqn} for a review) of $D=11$ \cite{Cremmer:1978km} and type II \cite{Giani:1984wc,Romans:1985tz,Schwarz:1983qr} supergravities are particularly helpful to investigate the imprint that these higher-dimensional theories instill on their lower-dimensional counterparts. In particular, Exceptional Field Theory (ExFT) \cite{Hohm:2013pua,Hohm:2013vpa,Hohm:2013uia} has been recently shown to provide a powerful framework to compute the spectrum of Kaluza-Klein (KK) perturbations above a certain class of anti-de Sitter (AdS) backgrounds of string and M-theory \cite{Malek:2019eaz,Malek:2020yue}. The relevant class of solutions involves the product, possibly warped, of AdS and a topologically-spherical manifold equipped with a possibly inhomogeneous metric, supported in general by fluxes. These solutions typically lie beyond the range of applicability of coset-space techniques \cite{Fabbri:1999mk,Ceresole:1999ht,Fre:1999gok,Nilsson:2018lof} for the calculation of KK spectra. Direct calculation methods \cite{Kim:1985ez} for these solutions become hard to the point of essentially unsuitable either. Using these novel ExFT techniques, the complete KK spectrum of some supersymmetric AdS solutions of the higher-dimensional supergravities in the relevant class has now been computed \cite{Malek:2019eaz,Malek:2020yue,Varela:2020wty,Eloy:2020uix}. The KK scalar \cite{Malek:2020mlk,Guarino:2020flh} and vector \cite{Varela:2020wty} spectrum of some non-supersymmetric AdS solutions in the same class has also been determined.

Specifically, the ExFT methods of \cite{Malek:2019eaz,Malek:2020yue} apply to $D=10, 11$ solutions with lower-dimensional AdS factors, which consistently uplift on spheres from AdS vacua of gauged supergravities in lower dimensions with certain gaugings. Under these conditions, infinite-dimensional mass matrices for each species of fields (scalars, vectors, etc.)~on AdS exist which are block-diagonal KK level by KK level. The consistent truncation requirement is critical for this block-diagonal structure, as the latter is absent for AdS solutions which do not uplift from a maximally supersymmetric lower-dimensional gauged supergravity \cite{Nilsson:2018lof,Cesaro:2020piw}. The infinite-\-dim\-en\-sio\-nal mass matrices for the KK bosonic perturbations above the relevant class of AdS solutions have been determined from ExFT in \cite{Malek:2019eaz,Malek:2020yue}. See also \cite{Dimmitt:2019qla} for an early derivation of a covariant mass matrix for the KK gravitons, and \cite{Varela:2020wty} for an alternative rederivation of the KK vector mass matrix from ExFT. In this paper, we complete this programme by providing the mass matrices for the fermionic, gravitino and spin-$1/2$, KK mass matrices. 

For definiteness, we will focus on E$_{7(7)}$ ExFT \cite{Hohm:2013uia,Ciceri:2016dmd}, and extract the KK fermion mass matrices from the fermionic completion of this theory \cite{Godazgar:2014nqa}. We thus focus on fermionic KK spectra above AdS$_4$ solutions, but our methods are readily extensible to other instances of ExFT with different duality groups. Also for concreteness, we will restrict our attention to the AdS$_4$ solutions of $D=11$ \cite{Cremmer:1978km} and massive type IIA supergravity \cite{Romans:1985tz} that uplift consistently on $S^7$  \cite{deWit:1986iy} and $S^6$ \cite{Guarino:2015jca,Guarino:2015vca} from AdS vacua of $D=4$ $\cN=8$ supergravity \cite{deWit:2007mt} with concrete gaugings. We take these to be, respectively, the SO(8) gauging \cite{deWit:1982ig} and the dyonic, in the sense of \cite{Dall'Agata:2012bb,Dall'Agata:2014ita,Inverso:2015viq}, ISO(7) gauging \cite{Guarino:2015qaa}. The latest classification results for this type of AdS$_4$ backgrounds of M-theory and type IIA string theory can be found in the recent references \cite{Comsa:2019rcz,Bobev:2020qev}. Section \ref{sec:KKFermionMassMat} below presents the KK fermionic mass matrices for this class of AdS$_4$ solutions.

%%%%%%%%%%%%%%%%%%%%%%%%%%%%%%%%%%%%%%%%%%%%%%%%%%%%%%%%%%%%%

\baselineskip=16pt
\begin{table}[]
\centering
\resizebox{\textwidth}{!}{
\begin{tabular}{l|c|ccccccccc}
\hline
Supermultiplet
& 	
$(s_0, E_0)$ 
&			
$s=2$
&
$s=\tfrac32$
&
$s=\tfrac32$
&
$s=1$
&
$s=1$
&
$s=\tfrac12$ 
&
$s=\tfrac12$ 
&
$s=0$&
$s=0$ \\
\hline\hline
\\[-10pt]
Massless graviton (MGRAV)
& 	
$(\tfrac32 , \tfrac52)$ 
&			
$3$
&
$\tfrac52$
&
$-$
&
$-$
&
$-$
&
$-$
&
$-$
&
$-$
&
$-$
\\[5pt]
Graviton (GRAV)
& 	
$(\tfrac32 , E_0)$ 
&			
$E_0 + \tfrac12$
&
$E_0 + 1$
&
$E_0 $
&
$E_0 + \tfrac12 $
&
$-$
&
$-$
&
$-$
&
$-$
&
$-$
\\[5pt]
Gravitino (GINO)
& 	
$(1 , E_0)$ 
&
$-$
&
$-$
&			
$E_0 + \tfrac12$
&
$E_0 + 1$
&
$E_0 $
&
$E_0 + \tfrac12 $
&
$-$
&
$-$
&
$-$
\\[5pt]
Massless vector (MVEC)
&
$(\tfrac12 , \tfrac32)$ 
&
$-$
&
$-$
&
$-$
&
$-$ 
&			
$2$
&
$\tfrac32$
&
$-$
&
$-$
&
$-$
\\[5pt]
Vector (VEC)
& 	
$( \tfrac12 , E_0)$ 
&
$-$
&
$-$
&
$-$
&
$-$
&			
$E_0 + \tfrac12$
&
$E_0 + 1$
&
$E_0 $
&
$E_0 + \tfrac12 $
&
$-$
\\[5pt]
Scalar (CHIRAL)
& 	
$(0 , E_0)$ 
&
$-$
&
$-$
&
$-$
&
$-$
&
$-$
&			
$-$
&
$E_0 + \tfrac12$
&
$E_0 + 1 $
&
$E_0  $
\\[5pt]  \hline
\end{tabular}
}
\caption{\footnotesize{OSp$(4|1)$ supermultiplets that appear in the $\cN=1$ KK spectra.  For each type of supermultiplet, the spin and energy $(s_0, E_0)$ of the superconformal primary  is given, and the energies of the constituent states, all of them with spins between 0 and 2, are listed.
}\normalsize}
\label{tab:OSp(4|1) supermultiplets}
\end{table}
%%%%%%%%%%%%%%%%%%%%%%%%%%%%%%%%%%%%%%%%%%%%%%%%%%%%%%%%%%%%%
%%%%%%%%%%%%%%%%%%%%%%%%%%%%%%%%%%%%%%%%%%%%%%%%%%%%%%%%%%%%%

We have then used our mass matrices to compute the KK gravitino and spin-$1/2$ spectra above concrete AdS$_4$ solutions in this class: those that preserve at least SU(3) internal symmetry. Prior to the general scans of \cite{Comsa:2019rcz,Bobev:2020qev}, these particular solutions were classified in \cite{Warner:1983vz} and \cite{Guarino:2015qaa} for the SO(8) and the ISO(7) gaugings, respectively. Their corresponding $D=11$ \cite{Freund:1980xh,Corrado:2001nv,deWit:1984nz,Godazgar:2013nma,deWit:1984va,Englert:1982vs,Pope:1984bd} and type IIA uplifts \cite{Guarino:2015jca,Guarino:2015vca,Varela:2015uca} are all known. For some of these AdS$_4$ solutions, the complete KK spectrum is also known \cite{Englert:1983rn,Sezgin:1983ik,Biran:1983iy,Klebanov:2008vq,Malek:2020yue,Varela:2020wty}, and we reproduce the corresponding fermionic sectors. More interestingly, there are three $\cN=1$ solutions in this class whose KK spectrum was only known partially \cite{Bobev:2010ib,Borghese:2012qm,Guarino:2015qaa,Dimmitt:2019qla,Pang:2017omp,Varela:2020wty} until now. One of these AdS$_4$ solutions has internal symmetry G$_2$  in the $D=4$ $\cN=8$ SO(8)  \cite{Warner:1983vz} and $D=11$ supergravities \cite{deWit:1984nz}. The other two are solutions of $D=4$ $\cN=8$ ISO(7) supergravity and type IIA. The first one of these also has residual symmetry G$_2$  in $D=4$ \cite{Borghese:2012qm} and type IIA \cite{Behrndt:2004km,Guarino:2015vca}. The second one has SU(3) symmetry in $D=4$ \cite{Borghese:2012zs,Guarino:2015qaa} and type IIA \cite{Varela:2015uca}. Combining our new fermionic spectra and previous partial results on the bosonic spectra \cite{Bobev:2010ib,Borghese:2012qm,Guarino:2015qaa,Dimmitt:2019qla,Pang:2017omp,Varela:2020wty}, we are able to give the complete supersymmetric spectra for all these three $\cN=1$ solutions. The details can be found in sections \ref{sec:N=1Spectrum11D} and \ref{sec:N=1Spectrum10D}.

Interestingly, a pattern emerges. The KK spectra for these three $\cN=1$ AdS$_4$ solutions are organised in representations of OSp$(4|1) \times G$, with $G= \textrm{G}_2$ or $G= \textrm{SU}(3)$. The supermultiplets of OSp$(4|1)$ \cite{Heidenreich:1982rz} have been reviewed for convenience in table \ref{tab:OSp(4|1) supermultiplets} above, and $G$, being of rank $2$, has its representations labelled by two non-negative integer Dynkin labels $[p,q]$. At Kaluza-Klein level $n=0 , 1, 2 , \ldots$, the dimension $E_0$ of a given OSp$(4|1)$ supermultiplet with superconformal primary spin $s_0$, arising in the $[p,q]$ representation of $G$, is found to be given by
\begin{equation} \label{eq:E0generic}
E_0 = 1 + \sqrt{ 6-s_0(s_0+1) + \alpha \, n (n + d-1) - \beta \, {\cal C}_2 (p,q) } \; .
\end{equation}
Here, $n( n + d-1)$ is the eigenvalue of the scalar Laplacian on the $S^{d}$ sphere, with $d=7$ in M-theory and $d=6$ in type IIA; $\alpha$ is a positive constant that takes on the value $\alpha = \tfrac58$ in M-theory and $\alpha = \tfrac56$ in type IIA for the specific $\cN=1$ solutions with  $G= \textrm{G}_2$ or $G= \textrm{SU}(3)$ symmetry; $\beta$ is a positive constant that depends on the symmetry preserved by the solution, $\beta = \tfrac54$ for  $G= \textrm{G}_2$ and $\beta = \tfrac53$ for  $G= \textrm{SU}(3)$, regardless of whether it lives in $D=11$ or in type IIA; and, finally, ${\cal C}_2 (p,q)$ is the eigenvalue of the quadratic Casimir operator of $G$ in the $[p,q]$ representation. 

Although we mainly focus on complete $\cN=1$ spectra, in section \ref{sec:N=0G2SpectrumIIA} we turn to give the fermionic spectra of the non-supersymmetric solutions in the same class. Of course, these $\cN=0$ solutions are of limited significance, since they are either manifestly unstable at the perturbative level \cite{Bobev:2010ib,DallAgata:2011aa} or expected to be so in the full string theory \cite{Ooguri:2016pdq}. These are either $D=11$ or type IIA solutions preserving $G = \textrm{SO}(7)$ \cite{deWit:1984va,Englert:1982vs,DallAgata:2011aa,Varela:2015uca,Romans:1985tz}, $G = \textrm{SO}(6) \sim \textrm{SU}(4)$ \cite{Pope:1984bd,DallAgata:2011aa,Varela:2015uca}, or $G = \textrm{G}_2$ \cite{Borghese:2012qm,Varela:2015uca,Lust:2008zd}. A formula (see (\ref{eq:M2L2non-susy})), similar to (\ref{eq:E0generic}) but now for the individual squared masses, exists in terms of the eigenvalues of the scalar Laplacian on $S^d$ and the quadratic Casimir operator of $G$. For the non-supersymmetric type IIA solution with G$_2$ invariance \cite{Borghese:2012qm,Varela:2015uca,Lust:2008zd}, our fermionic results combined with the previously known bosonic KK sector \cite{Borghese:2012qm,Pang:2017omp,Varela:2020wty,Guarino:2020flh} allow us to give its complete KK spectrum. For the other non-supersymmetric solutions, the only sector of the KK spectrum that remains to be explicitly computed after our analysis is the scalar sector. However, the pattern displayed by the generic mass formula (\ref{eq:M2L2non-susy}) is sufficiently strong to allow us to confidently conjecture the form of the KK scalar spectra for these solutions. 

Section \ref{sec:Discussion} concludes with further discussion. Our conventions are summarised in appendix \ref{sec:SpecificSpectra}, where some explicit results for the eigenvalues of our fermionic mass matrices on selected solutions are also included.

%%%%%%%%%%%%%%%
%%%%%%%%%%%%%%%

\section{KK fermion mass matrices from ExFT} \label{sec:KKFermionMassMat}

%%%%%%%%%%%%%%%
%%%%%%%%%%%%%%%

We will now determine the mass matrices for the KK fermion perturbations above the AdS$_4$ class of solutions of string and M-theory that uplift from four-dimensional gauged supergravity. We will extract these mass matrices from the fermionic completion \cite{Godazgar:2014nqa} of E$_{7(7)}$ ExFT \cite{Hohm:2013uia,Ciceri:2016dmd}, by setting the ExFT bosonic fields to the Scherk-Schwarz configuration that gives rise to $D=4$ $\cN=8$ supergravity upon consistent truncation, while retaining the full tower of KK fermion perturbations.

%%%%%%%%%%%%%%%

\subsection{Generalised Scherk-Schwarz--Kaluza-Klein reduction} \label{sec:KKVecPrelim}

The fermionic content of ExFT includes a gravitino $\bm{\psi}_\mu^i$ and a spin-$1/2$ fermion $\bm{\chi}^{ijk} = \bm{\chi}^{[ijk]}$, neutral under local E$_{7(7)}$ but transforming in the $\bm{8}$ and the $\bm{56}$ of global SU(8), respectively. The bosonic sector of ExFT contains external, $\bm{e}_\mu{}^\alpha$, and internal, $\bm{ \mathcal{V}}_M{}^{\underline{A}} = \big( \bm{ \mathcal{V}}_M{}^{ij} , \bm{ \mathcal{V}}_{M \,ij} \big)$, vielbeine which give rise to metrics $\bm{g}_{\mu \nu} = \eta_{\alpha \beta} \, \bm{e}_\mu{}^\alpha \bm{e}_\nu{}^\beta$ and $\bm{\mathcal{M}}_{MN}  = 2 \, \bm{ \mathcal{V}}_{(M |  \,ij} \bm{ \mathcal{V}}_{|N)}{}^{ij} $. The indices $\mu = 0, \ldots ,  3$ and $M=1 , \ldots , 56$ are local fundamental indices of SO$(1,3)$ and E$_{7(7)}$, while $\underline{A} = \big( \phantom{}^{ij} ,  \phantom{}_{ij}  \big) $ and $i$ are global indices in the $\bm{28} + \overline{\bm{28}}$ and the fundamental of SU(8), respectively, so that $\bm{ \mathcal{V}}_{M \,ij} = \big( \bm{ \mathcal{V}}_M{}^{ij} \big)^*$ and $\bm{ \mathcal{V}}_M{}^{ij} = \bm{ \mathcal{V}}_M{}^{[ij]} $. The bosonic sector of E$_{7(7)}$ ExFT further includes vectors and two-forms which will not play a role in the present analysis. All these fields depend on both the external and internal coordinates $(x^\mu,Y^M)$, and are subject to the appropriate section constraints.

Our starting point is the fermionic action of ExFT \cite{Godazgar:2014nqa}. The terms that contribute to the kinetic, mass, and quadratic interaction terms for the $D=4$ KK fermions are, in our conventions \cite{Godazgar:2014nqa} 
{\setlength\arraycolsep{2pt}
\begin{eqnarray}\label{eq:fermiLagrangian}
\mathcal{L}_{\text{ExFT fermi}} &=& - i \epsilon^{\mu\nu \rho\sigma}\bar{\bm{\psi}}_\mu^i \gamma_\nu \mathcal{D}_\rho \bm{\psi}_{\sigma i} 
- \tfrac{1}{6}\, \bm{e} \,\bar{\bm{\chi}}^{ijk} \gamma^\mu \mathcal{D}_\mu \bm{\chi}_{ijk} \nonumber \\[4pt]
& & -4 i  \, \epsilon^{\mu\nu \rho\sigma}  \big( \bm{ {\cal V}}^{-1} \big)_{ij}{}^M \, \bar{\bm{\psi}}_\mu^i \gamma_\nu \nabla_M\left(\gamma_\rho \bm{\psi}_\sigma^j\right)
- 4 \sqrt{2} \, \bm{e} \,\big( \bm{ {\cal V}}^{-1}\big)^{ij \, M} \, \bar{\bm{\psi}}_\mu^k \, \nabla_M \left(\gamma^\mu  \bm{\chi}_{ijk}\right) \nonumber \\[4pt]
&& + \tfrac{1}{9}\, \bm{e} \, \epsilon_{ijklmnpq } \big( \bm{ {\cal V}}^{-1}\big)^{ij \, M}  \bar{\bm{\chi}}^{klm} \nabla_M \bm{\chi}^{npq} \, +\, \text{c.c.} 
\end{eqnarray}
}Here, $\bm{e} \equiv \sqrt{\lvert \text{det} \, \bm{g}_{\mu\nu} \rvert}$, $\gamma_\alpha$ are the SO$(1,3)$ Dirac matrices subject to the Clifford algebra $\{ \gamma_\alpha , \gamma_\beta \} = 2 \eta_{\alpha \beta}$, and $\gamma_\mu \equiv \bm{e}_\mu{}^\alpha \, \gamma_\alpha$. The external covariant derivatives featuring in the kinetic terms (the first two terms in (\ref{eq:fermiLagrangian})) take on the schematic form $\mathcal{D}_\mu \equiv D_\mu+ \tfrac14 \omega_{\mu\, \;\beta}^{\;\;\alpha}+ \tfrac12 \mathcal{Q}_{\mu \,\;j}^{\;\;i}$, in terms of SO(1,3) and SU(8) connections, $\omega_{\mu\, \;\beta}^{\;\;\alpha}$ and $\mathcal{Q}_{\mu \,\;j}^{\;\;i}$. The portion $D_\mu \equiv \partial_\mu - \mathbb{L}_{\mathcal{A}_\mu}$, covariant under generalised diffeomorphisms generated by the ExFT gauge fields, will not be significant. The second and third lines in (\ref{eq:fermiLagrangian}) contain the internal covariant derivative $\nabla_M$. On an SU(8) vector $\bm{\xi}_i$, with suppressed SO(1,3) spinor indices and weight $\lambda$ under generalised diffeomorphisms, this derivative acts as \cite{Godazgar:2014nqa}
\begin{equation} \label{eq:CovInt}
\nabla_M \bm{\xi}_i = \partial_M \bm{\xi}_i -\tfrac14 \omega_M{}^{\alpha \beta} \gamma_{\alpha \beta} \, \bm{\xi}_i + \tfrac12 {\cal Q}_{M \, i}{}^j \bm{\xi}_j -\tfrac23 \, \lambda \, \Gamma_{KM}{}^K \bm{\xi}_i \; , 
\end{equation}
in terms of internal SO$(1,3)$, SU(8) and Christoffel connections $\omega_M{}^{\alpha \beta} \equiv \bm{e}^{\mu [\alpha}\partial_M \bm{e}_\mu{}^{\beta]}$, ${\cal Q}_{M \, i}{}^j $ and $\Gamma_{MN}{}^P$. There are other terms in the quadratic fermionic action \cite{Godazgar:2014nqa} that we have not retained in (\ref{eq:fermiLagrangian}), since they do not contribute to the mass matrices. For example, there is a coupling of the gravitino and the spin-$1/2$ fermion to the spacetime Maurer-Cartan form of the scalars. For similar reasons, a contribution to the SO$(1,3)$ connection $\omega_M{}^{\alpha \beta}$ containing $\bm{\mathcal{M}}_{MN}$ together with the ExFT gauge field strengths has been disregarded. See \cite{Godazgar:2014nqa} for full details.

We would now like to fix the external and internal vielbeine to the configurations that give rise to $D=4$ $\cN=8$ gauged supergravity upon consistent truncation. We therefore write for them the generalised Scherk-Schwarz expressions
\cite{Berman:2012uy,Lee:2014mla,Hohm:2014qga,Cassani:2016ncu} 
\begin{equation} \label{eq:ScalarMetExFT}
\bm{e}_\mu{}^\alpha (x,Y)  = \rho(Y)^{-1} \, e_\mu{}^\alpha (x) \; , \qquad
\bm{ \mathcal{V}}_M{}^{\underline{A}}  (x, Y) = U_{M}{}^{\underline{M}} (Y) \, \mathcal{V}_{\underline{M}}{}^{\underline{A}} (x)  \; ,
\end{equation}
in terms of the $D=4$ vielbein $e_\mu{}^\alpha (x)$ and $\textrm{E}_{7(7)}/\textrm{SU}(8)$ coset representative $\mathcal{V}_{\underline{M}}{}^{\underline{A}} (x)$. The $Y$-dependent function $\rho$ and twist matrix $U_{M}{}^{\underline{M}} (Y) $ here obey the consistency conditions
 \cite{Lee:2014mla,Hohm:2014qga,Cassani:2016ncu,Inverso:2017lrz} 
\begin{eqnarray} \label{eq:ConsistentKK}
&& \partial_N \, (U^{-1})_{\underline{M}}{}^N - 3 \rho^{-1}  (U^{-1})_{\underline{M}}{}^N\partial_N \, \rho =0 \; , \nonumber \\[5pt]
&& 7 \rho^{-1} \, \left( (U^{-1})_{\underline{M}}{}^P \, (U^{-1})_{\underline{N}}{}^Q \, \partial_P \, U_Q{}^{\underline{K} }\right)_{\bm{912}} + F_{\underline{M} \underline{N}}{}^{\underline{K}}  =  X_{\underline{M} \underline{N}}{}^{\underline{K}} \; ,
\end{eqnarray}
where $X_{\underline{M} \underline{N}}{}^{\underline{P}} \equiv \Theta_{\underline{M}}{}^\alpha \, (t_\alpha)_{\underline{N}}{}^{\underline{P}} $ is the usual contraction of the $D=4$ $\cN=8$ embedding tensor and the E$_{7(7)}$ generators, the subindex $\bm{912}$ denotes projection to that representation of E$_{7(7)}$, and $F_{\underline{M} \underline{N}}{}^{\underline{P}}$ is a deformation \cite{Ciceri:2016dmd,Inverso:2017lrz} of E$_{7(7)}$ ExFT \cite{Hohm:2013uia}. The latter either vanishes or codifies the Romans mass for $D=11$ and type IIA configurations, respectively. In (\ref{eq:ScalarMetExFT}) and elsewhere, $\underline{M} = 1, \ldots , 56$ is a flat, fundamental $\textrm{E}_{7(7)}$ index.

We would also like to keep the full tower of KK gravitini and spin-$1/2$ fermion perturbations over every AdS vacuum of 
the $D=4$ $\cN=8$ supergravities under consideration. Identifying, in our conventions, the $\bm{8}$ of SU(8) with the $\bm{8}_s$ of SO(8), the $D=4$ $\cN=8$ gravitino $\psi_\mu^i$ and spin-$1/2$ fermion $\chi^{ijk}$ respectively lie in the $\bm{8}_s \equiv [0,0,0,1]$ and $\bm{56}_s \equiv [1,0,1,0]$ of SO(8) for AdS vacua of the SO(8) gauging (or branchings thereof for vacua with reduced symmetry $G \subset \textrm{SO}(8)$), and in the $\bm{8} \equiv [0,0,1]$ and $\bm{48} + \bm{8} \equiv [1,0,1] + [0,0,1]$ of SO(7) (or branchings thereof) for vacua of the ISO(7) gauging. These $D=4$ $\cN=8$ states are identified with the KK level $n=0$ states. The $n \geq 1$ states up the KK tower lie in the infinite-dimensional, reducible representation obtained by tensoring the representations above with the symmetric-traceless representations \cite{Englert:1983rn,Malek:2019eaz,Varela:2020wty}
\begin{equation} \label{eq:SymTrac}
\oplus_{n=0}^\infty [n,0,0,0] \; \textrm{of SO(8)} \quad \textrm{or} \quad   \oplus_{n=0}^\infty  [n,0,0] \; \textrm{of SO(7)} \; ,
\end{equation}
at least, for the spin-$1/2$ fermions, before super-Higgsing takes place: see the discussion around (\ref{eq:IrrepsA1KKSO8Spin1/2}) and (\ref{eq:IrrepsA1KKISO7Spin1/2}) below. It is thus convenient to denote the KK gravitino and spin-$1/2$ fermions with a double set of indices carrying this tensor product structure, as $\psi_\mu^{i \Lambda} (x)$ and $\chi^{ijk \Lambda} (x) $. 

For the ExFT fermions we thus write
\begin{equation}\label{eq:SSpertfermions}
\bm{\psi} _\mu^i\left(x,Y\right) = \rho (Y )^{-\frac{1}{2}}  \,\psi_\mu^{i  \Lambda}(x)\, \mathcal{Y}_\Lambda (Y ) \, , \qquad 
\bm{\chi}^{ijk}\left(x,Y\right)  = \rho (Y )^{\frac{1}{2}} \,\chi^{ijk  \Lambda}(x)\,\mathcal{Y}_\Lambda  (Y )  \, ,
\end{equation}
building on \cite{Hohm:2014qga,Malek:2019eaz}. Here, ${\cal Y}_\Lambda$ denotes the infinite tower of scalar spherical harmonics on the round $S^7$ or $S^6$ spheres. These lie in the representations (\ref{eq:SymTrac}) of SO(8) or SO(7). Of the ${\cal Y}_\Lambda$ we will only need to note that they are subject to the relation \cite{Malek:2019eaz,Malek:2020yue}
\begin{equation} \label{eq:ActionSH}
\rho^{-1} \, (U^{-1})_{\underline{N}}{}^M \partial_M \, {\cal Y}_\Lambda = -({\cal T}_{\underline{N}})_\Lambda{}^\Sigma \, {\cal Y}_\Sigma \; ,
\end{equation}
where the constant, real matrices  $({\cal T}_{\underline{N}})_\Lambda{}^\Sigma$ are the generators of SO(8) or SO(7) in the infinite-dimensional, reducible representations (\ref{eq:SymTrac}), normalised as \cite{Malek:2019eaz,Malek:2020yue}
\begin{equation} \label{eq:SOGen}
[ {\cal T}_{\underline{M}} , {\cal T}_{\underline{N}} ] = -X_{\underline{M} \underline{N}}{}^{\underline{P}} \, {\cal T}_{\underline{P}} \; .
\end{equation}
These are of course traceless, $({\cal T}_{\underline{N}})_\Lambda{}^\Lambda =0$. Indices $\Lambda$, $\Sigma$ are raised and lowered with $\delta_{\Sigma \Lambda}$, and the generators with same-level indices are antisymmetric,
\begin{equation} \label{eq:Tantisym}
({\cal T}_{\underline{M}})_{\Lambda \Sigma} \equiv ({\cal T}_{\underline{M}})_\Lambda{}^\Xi \,  \delta_{\Xi \Sigma}  = - ({\cal T}_{\underline{M}})_{\Sigma \Lambda } \; ,\qquad
({\cal T}_{\underline{M}})^{\Lambda \Sigma} \equiv \delta^{\Lambda \Xi} ({\cal T}_{\underline{M}})_\Xi{}^\Sigma  = - ({\cal T}_{\underline{M}})^{\Sigma \Lambda } \; .
\end{equation}
At KK level $n=0$, ${\cal Y}_0 = 1$ and (\ref{eq:SSpertfermions}) reduces to the expressions given in \cite{Hohm:2014qga} for the consistent truncation of the ExFT fermions down to their $D=4$ $\cN=8$ counterparts. For higher KK levels $n \geq 1$, (\ref{eq:SSpertfermions}) is a straightforward extension to the fermion sector of the expressions given in \cite{Malek:2019eaz,Malek:2020yue} for the embedding of the bosonic KK modes into the ExFT bosonic fields.

Finally, we still need to give expressions for the ExFT connections in terms of $D=4$ quantities. With the ExFT scalars fixed via the rightmost equation in (\ref{eq:ScalarMetExFT}) to their $\cN=8$ four-dimensional counterparts (and the latter eventually frozen to their vacuum expectation values at an AdS critical point of the $\cN=8$ scalar potential), the ExFT connections simply take on their expressions for consistent truncation configurations \cite{Hohm:2014qga}. In particular, the internal SO$(1,3)$ connection $\omega_M{}^{\alpha \beta}$ and the relevant components, ${\cal Q}^{ik}{}_k{}^j$ and ${\cal Q}^{[ij}{}_k{}^{l]}$, of the flattened SU(8) connection ${\cal Q}^{ij}{}_k{}^{l} \equiv \big( \bm{ {\cal V}}^{-1}\big)^{ij \, M } \, {\cal Q}_{M \, k}{}^l$ are set, in our conventions, to \cite{Hohm:2014qga}
\begin{equation} \label{eq:SU8connectionSS}
\omega_M{}^{\alpha \beta} (x, Y) = 0 \; , \qquad
 \mathcal{Q}^{ik}{}_k{}^j =- \tfrac14 \,  \rho(Y) \,  A_1^{ij} (x)  \; , \qquad
{\cal Q}^{[ij}{}_k{}^{l]} =- \tfrac{1}{12} \, \rho(Y) \,  A_{2 \, k}{}^{ijl} (x) \; .
\end{equation}
Here, $A_1^{ij}$ and  $A_{2 \, k}{}^{ijl}$ are the fermion shifts of $\cN=8$ gauged supergravity. Recall that these arise as contractions \cite{deWit:2007mt}
\begin{equation}\label{eq:A1A2}
 A_1^{ij} =\tfrac{4}{21} \, T^{ikjl}{}_{kl} \quad , \quad A_{2h}{}^{ijk} = 2 \, T_{mh}{}^{mijk}\, ,
 \end{equation}
of the $D=4$ $\cN=8$ $T$-tensor \cite{deWit:2007mt}
\begin{equation}\label{eq:flatembtensor}
T_{\underline{A}\underline{B}}{}^{\underline{C}}=\left(\mathcal{V}^{-1}\right)_{\underline{A}}{}^{\underline{M}} \left(\mathcal{V}^{-1}\right)_{\underline{B}}{}^{\underline{N}} \, X_{ \underline{M} \underline{N}}{}^{\underline{P}} \,  \mathcal{V}_{\underline{P}}{}^{\underline{C}} \; .
\end{equation}
More specifically, $A_1^{ij}$ and $A_{2 \, k}{}^{ijl}$ are the $\bm{36}$ and $\bm{420}$ components of the $T$-tensor (\ref{eq:flatembtensor}), so that
\begin{eqnarray} \label{eq:A1A2sym}
A_{1}^{ ij} = A_{ 1}^{ (ij)} \;  , \quad A_{2 h}{}^{ijk} = A_{2  h}{}^{[ijk]} \; , \quad A_{2  k}{}^{ijk} = 0 \; . 
\end{eqnarray}
These are related to the $\overline{\bm{36}}$ and $\overline{\bm{420}}$ components $A_{1 ij}$, $A_2{}^h{}_{ijk}$  of (\ref{eq:flatembtensor}) by complex conjugation: $( A_{1 ij})^* =  A_1^{ij}$, $(A_2{}^h{}_{ijk})^* = A_{2h}{}^{ijk}$. The $\cN=8$ fermion shift $A_{1 ij}$ serves also as the gravitino mass matrix, while the $\cN=8$ spin-$1/2$ fermion mass matrix $A_{3 \, ijk , lmn}$ is 
symmetric,
\begin{equation} \label{eq:A3Sym}
A_{3 \, ijk , lmn} =A_{3 \,  lmn,  ijk} \; , 
\end{equation}
and related to $A_{2 h}{}^{ijk}$ via\footnote{The relation between $A_2$ and $A_3$ is usually given as \cite{deWit:2007mt}
\begin{equation} \label{eq:A3A2bis}
A_{3 \, ijk , lmn} = \tfrac{\sqrt{2}}{144} \, \epsilon_{ijkpqr[lm} \,  A_{2 \, n]}{}^{pqr} \; ,
\end{equation} 
which, unlike (\ref{eq:A3A2}), makes manifest the symmetry (\ref{eq:A3Sym}) of $A_3$. Using the tracelessness condition of $A_2$ in (\ref{eq:A1A2sym}), the expression (\ref{eq:A3A2}) becomes equivalent to (\ref{eq:A3A2bis}). The KK analogue of $A_2$ satisfies a similar trace condition only if the KK indices are contracted as well. For this reason, it is (\ref{eq:A3A2}) and not (\ref{eq:A3A2bis}) that naturally extrapolates to higher KK levels: see (\ref{eq:A1A2A3KK})}
\begin{equation} \label{eq:A3A2}
A_{3 \, ijk , lmn}  =  \tfrac{\sqrt{2}}{108} \, \epsilon_{ijkpq[rlm} \,  A_{2 n]}{}^{pqr} \; .
\end{equation}

Armed with all these definitions, our task now is to obtain from (\ref{eq:fermiLagrangian}) the $D=4$ quadratic action for the KK fermion pertubations $\psi^{i \Lambda} (x)$ and  $\chi^{ijk \Lambda} (x) $. The calculation follows closely that described in \cite{Hohm:2014qga} to obtain the supersymmetry variations of the $D=4$ $\cN=8$ fermions from ExFT. One should now take into account the dependence (\ref{eq:SSpertfermions}) of the ExFT fermions on the spherical harmonics ${\cal Y}_\Lambda$. On top of the $\cN=8$ contributions \cite{Hohm:2014qga}, these amount to new terms in the generators $({\cal T}_{\underline{N}})_\Lambda{}^\Sigma$. As will be shown in sections \ref{sec:KKGINOMassMatCalc} and \ref{sec:KKSpin1/2MassMatCalc}, the end result is
{\setlength\arraycolsep{2pt}
\begin{eqnarray} \label{eq:KKFermionLag}
{\cal L}_{\text{KK  fermi}} &=& -i \epsilon^{\mu\nu \rho\sigma}\bar{\psi}_{\mu}^{i  \Lambda}\, \gamma_\nu \,\mathcal{D}_\rho \psi_{\sigma i  \Lambda}
-\tfrac{1}{6}\,e\,\bar{\chi}^{ijk \Lambda} \, \gamma^\mu \, \mathcal{D}_\mu \chi_{ijk  \Lambda} \nonumber \\[4pt]
&& +  e \, A_{1 \, i \Lambda , j \Sigma} \; \bar{\psi}_\mu^{i \Lambda}\,\gamma^{\mu \nu} \, \psi_{\nu}^{j \Sigma}
+ \tfrac{\sqrt{2}}{6} \,e \,  A_{2 \, i \Lambda}{}^{jkl \Sigma} \; \bar{\psi}_{\mu}^{i \Lambda}\, \gamma^\mu \,\chi_{jkl \Sigma}
 \nonumber \\[4pt]
&& + \sqrt{2} \, e \, A_3^{ijk \Lambda , lmn \Sigma}  \, \bar{\chi}_{ijk \Lambda }\, \chi_{lmn \Sigma}  \; + \, \text{c.c.} 
\end{eqnarray}
}External indices are now raised and lowered with the $D=4$ metric $g_{\mu \nu} = \eta_{\alpha \beta} \,  e_\mu{}^\alpha e_\nu{}^\beta$, and $e \equiv \sqrt{\lvert \text{det} \, g_{\mu\nu} \rvert}$. 

The external covariant derivatives in the kinetic terms of the $D=4$ action (\ref{eq:KKFermionLag}) reduce as in \cite{Hohm:2014qga}, now also including new couplings to the $D=4$ $\cN=8$ gauge fields $A_\mu^{\underline{M}} (x)$, had we retained them, of the form $A_\mu^{\underline{M}} \, ({\cal T}_{\underline{M}})_\Lambda{}^\Sigma \, \psi_{\nu i \Sigma}$. More importantly for our purposes, the gravitino and spin-$1/2$ mass terms and their quadratic interactions in (\ref{eq:KKFermionLag}) are codified by new tensors $A_{1 \, i \Lambda , j \Sigma}$, $A_{2 \, i \Lambda}{}^{jkl \Sigma}$, $A_3{}^{ijk \Lambda , lmn \Sigma}$. 
These reduce to their gauged supergravity counterparts $A_{1 i j}$, $A_{2i }{}^{jkl }$, $A_3^{ijk , lmn }$ at KK level $n=0$ and extrapolate them at higher KK levels. Like their $n=0$ versions, the new tensors $A_{1 \, i \Lambda , j \Sigma}$, $A_{2 \, i \Lambda}{}^{jkl \Sigma}$, $A_3{}^{ijk \Lambda , lmn \Sigma}$ depend on the $D=4$ $\cN=8$ scalars. Further, they obey
\begin{eqnarray} \label{eq:A1A2A3KK}
& A_{1 \, i \Lambda , j \Sigma} =  A_{1 \, j \Sigma ,  i \Lambda }  \; , \qquad
A_{2 \, h  \Lambda }{}^{ijk \Sigma} = A_{2 \, h  \Lambda }{}^{[ijk] \Sigma}  \; , \qquad
A_{2 \, k  \Lambda }{}^{ijk \Lambda} = 0 \;  , \nonumber \\[5pt]
& A_{3 \, ijk\Lambda  , lmn \Sigma} = A_{3 \, lmn \Sigma , ijk\Lambda }  \; , \qquad
A_{3 \, ijk\Lambda  , lmn \Sigma} =  \tfrac{\sqrt{2}}{108} \, \epsilon_{ijkpq[rlm} \,  A_{2 \, n ] \Sigma}{}^{pqr \Omega } \, \delta_{\Omega \Lambda} \; , 
\end{eqnarray}
in direct analogy with the relations (\ref{eq:A1A2sym}), (\ref{eq:A3Sym}) and (\ref{eq:A3A2}) satisfied by their $D=4$ $\cN=8$ counterparts.

%%%%%%%%%%%%%%%

\subsection{The KK gravitino mass matrix} \label{sec:KKGINOMassMatCalc}

%%%%%%%%%%%%%%%

The KK gravitino mass term in the $D=4$ action (\ref{eq:KKFermionLag}) derives from the following piece in the ExFT action (\ref{eq:fermiLagrangian}) 
\begin{equation} \label{eq:massterm3/2}
\mathcal{L}_{ \bm{\psi} \bm{\psi}} =  - 4 i  \epsilon^{\mu\nu \rho\sigma}  \big( \bm{ {\cal V}}^{-1} \big)_{ij}{}^M \, \bar{\bm{\psi}}_\mu^i \gamma_\nu \nabla_M\left(\gamma_\rho \bm{\psi}_\sigma^j\right)  \; .
\end{equation}
Taking into account the form of the covariant derivative (\ref{eq:CovInt}) with weight $\lambda  \big(\gamma_\rho \bm{\psi}_\sigma^j \big) = \tfrac34$ \cite{Godazgar:2014nqa}, using the generalised Scherk-Schwarz expressions (\ref{eq:ScalarMetExFT}), (\ref{eq:SSpertfermions}), (\ref{eq:SU8connectionSS}) and getting rid of the Levi-Civita tensor (through $\gamma^{\mu\nu} = \frac{i}{2} \, e^{-1} \,  \epsilon^{\mu\nu\rho\sigma} \gamma_{\rho \sigma} \gamma_5$ and the chirality of the KK gravitini, $\gamma_5 \, \psi_\mu^{i\Lambda} = \psi_\mu^{i\Lambda}$), (\ref{eq:massterm3/2}) becomes
\begin{equation} \label{eq:massterm3/2SS}
\mathcal{L}_{ \bm{\psi} \bm{\psi}} =  e\left( \rho^{-2} A_{1 ij} \, \mathcal{Y}_\Lambda  \mathcal{Y}_\Sigma - 8 \,  \rho^{-3} \,( {\cal V}^{-1} )_{ij}{}^{\underline{N}}  (U^{-1})_{\underline{N}}{}^M {\cal Y}_\Lambda \partial_M \, {\cal Y}_\Sigma \right) \, \bar{\psi}_\mu^{i\Lambda} \,  \gamma^{\mu\nu} \psi_\nu^{j\Sigma} \; .
\end{equation}
The second term in the parenthesis can be further simplified using (\ref{eq:ActionSH}). After this substitution, (\ref{eq:massterm3/2SS}) scales with an overall factor of $\rho^{-2}$, as in fact does the entire ExFT action upon generalised Scherk-Schwarz reduction. This factor thus drops at the level of the equations of motion. Also, (\ref{eq:massterm3/2SS})  acquires a quadratic dependence on $\mathcal{Y}_\Lambda  \mathcal{Y}_\Sigma$ after (\ref{eq:ActionSH}) is used. This dependence reduces to $\delta_{\Lambda \Sigma}$ at the level of the action, by virtue of the orthogonality of the spherical harmonics. 

Thus, (\ref{eq:massterm3/2}) gives rise to the KK gravitino mass term in the $D=4$ action (\ref{eq:KKFermionLag}), with mass matrix
\begin{equation} \label{eq:A1KKdd}
 A_{1 \, i \Lambda , j \Sigma} \equiv A_{1 ij} \, \delta_{\Lambda \Sigma} - 8 \, ( {\cal V}^{-1} )_{ij}{}^{ \underline{M} }  ({\cal T}_{\underline{M}})_{\Lambda \Sigma} \; .
\end{equation}
Due to the symmetry of $A_{1 ij}$ and the antisymmetry of $( {\cal V}^{-1} )_{ij}{}^{ \underline{M} }$ and $({\cal T}_{\underline{M}})_{\Lambda \Sigma}$ (in the indices $ij$ and $\Lambda \Sigma$, respectively), the tensor $ A_{1 \, i \Lambda ,  j \Sigma}$ defined in (\ref{eq:A1KKdd}) is symmetric  in the product index $i\Lambda$, as it must from the action (\ref{eq:KKFermionLag}) and asserted in the first relation of (\ref{eq:A1A2A3KK}). Likewise, the complex conjugate (c.c.)~contribution of (\ref{eq:massterm3/2}) to the ExFT action (\ref{eq:fermiLagrangian}) reduces to a c.c.~contribution to the $D=4$ action (\ref{eq:KKFermionLag}) with
\begin{equation} \label{eq:A1KKuu}
 A_1{}^{i \Lambda , j \Sigma} \equiv A_{1}^{ ij} \, \delta^{\Lambda \Sigma} - 8 \, ( {\cal V}^{-1} )^{ ij \underline{M} }  ({\cal T}_{\underline{M}})^{\Lambda \Sigma} \; ,
\end{equation}
so that $ \big( A_{1 \, i \Lambda , j \Sigma} \big)^* =   A_1{}^{i \Lambda , j \Sigma}$ manifestly.

By (\ref{eq:SymTrac}), the KK gravitino mass matrix (\ref{eq:A1KKdd}) is an infinite-dimensional, block-diagonal square matrix. With the conventions specified above (\ref{eq:SymTrac}), for $D=11$ AdS$_4$ vacua that uplift from the SO(8) gauging, the square block at level $n$ comes in the SO(8) \mbox{representations}
\begin{equation} \label{eq:IrrepsA1KKSO8}
[0,0,0,1] \times [n,0,0,0] \, \longrightarrow \, [n,0,0,1]  + [n-1,0,1,0]  \; , 
\end{equation}
or their branchings thereof under $G \subset \textrm{SO}(8)$ for AdS$_4$ solutions with residual symmetry group $G$. For AdS$_4$ solutions of type IIA that uplift from the ISO(7) gauging, the block at KK level $n$ comes instead in the SO(7) representations
\begin{equation} \label{eq:IrrepsA1KKISO7}
[0,0,1]  \times [n,0,0] \, \longrightarrow \, [n,0,1]+[n-1,0,1] \; ,
\end{equation}
or their branchings under $G \subset \textrm{SO}(7)$ for AdS$_4$ solutions with symmetry group $G$. In (\ref{eq:IrrepsA1KKSO8}) and (\ref{eq:IrrepsA1KKISO7}) only representations with positive Dynkin labels must be kept. In both cases, all eigenvalues of the mass matrix (\ref{eq:A1KKdd}) are physical and contribute to the spectrum of physical KK gravitini.

%%%%%%%%%%%%%%%

\subsection{The KK spin-$1/2$ fermion mass matrix} \label{sec:KKSpin1/2MassMatCalc}

%%%%%%%%%%%%%%%

Following similar steps, we can determine the contributions of the gravitino--spin-$1/2$ fermion and fermion-fermion terms from the ExFT quadratic action (\ref{eq:fermiLagrangian})
\begin{equation}
{\cal L}_{\bm{\psi} \bm{\chi}} = -4\sqrt{2} \, \bm{e} \,\big( \bm{ {\cal V}}^{-1}\big)^{ij \, M} \, \bar{\bm{\psi}}_\mu^k \, \nabla_M \left(\gamma^\mu  \bm{\chi}_{ijk}\right)  , \quad
{\cal L}_{\bm{\chi} \bm{\chi}} = \tfrac{1}{9}\, \bm{e} \, \epsilon_{ijklmnpq } \big( \bm{ {\cal V}}^{-1}\big)^{ij \, M}  \bar{\bm{\chi}}^{klm} \nabla_M \bm{\chi}^{npq} .
\end{equation}
The ${\cal L}_{\bm{\psi} \bm{\chi}}$ term and its complex conjugate give rise to the corresponding KK gravitino-fermion terms in the $D=4$ KK action (\ref{eq:KKFermionLag}), with a KK tensor $A_2$ defined as
\begin{equation} \label{eq:A2KKduuu}
A_{2 \, i \Lambda}{}^{jkl \Sigma} \equiv A_{2i }{}^{jkl } \, \delta_\Lambda^\Sigma - 24 \,  \delta_i^{[j } \,  ( {\cal V}^{-1} )^{kl ] \underline{N} }  ({\cal T}_{\underline{N}})_{\Lambda}{}^\Sigma \; , 
\end{equation}
and
\begin{equation} \label{eq:A2KKuddd}
A_2{}^{i \Lambda}{}_{jkl \Sigma} \equiv A_2{}^{i }{}_{jkl } \, \delta^\Lambda_\Sigma - 24 \,  \delta^i_{[j } \,  ( {\cal V}^{-1} )_{kl ]}{}^{\underline{N} }  ({\cal T}_{\underline{N}})^{\Lambda}{}_\Sigma \; , 
\end{equation}
so that $(A_{2i \Lambda}{}^{jkl \Sigma} )^* = A_2{}^{i \Lambda}{}_{jkl \Sigma}$. The tensor $A_2$ given in (\ref{eq:A2KKduuu}) manifestly satisfies the antisymmmetry and tracelessness relations stated in (\ref{eq:A1A2A3KK}). 

Finally, the term ${\cal L}_{\bm{\chi} \bm{\chi}}$ in the ExFT action gives rise to the mass terms for the spin-$1/2$ fields in the $D=4$ KK action (\ref{eq:KKFermionLag}) with mass matrix proportional to
\begin{equation} \label{eq:A3dddddd}
A_{3 \,  ijk \Lambda , lmn \Sigma}  \equiv A_{3 \,  ijk , lmn} \, \delta_{\Lambda \Sigma} + \tfrac{\sqrt{2}}{18} \,  \epsilon_{ijklmnpq} \, {\cal} ( {\cal V}^{-1} )^{pq \underline{N} }  ({\cal T}_{\underline{N}})_{\Lambda \Sigma} \; , 
\end{equation}
along with its complex conjugate,
\begin{equation} \label{eq:A3uuuuuu}
A_{3}{}^{ ijk \Lambda , lmn \Sigma}  \equiv A_{3}{}^{ ijk , lmn} \, \delta^{\Lambda \Sigma} + \tfrac{\sqrt{2}}{18} \,  \epsilon^{ijklmnpq} \, {\cal} ( {\cal V}^{-1} )_{pq}{}^{ \underline{N} }  ({\cal T}_{\underline{N}})^{\Lambda \Sigma} \; , 
\end{equation}
so that $(A_{3 \,  ijk \Lambda , lmn \Sigma})^* = A_{3}{}^{ ijk \Lambda , lmn \Sigma}$. The tensor (\ref{eq:A3dddddd}) is manifestly symmetric in its product indices, as required by the $D=4$ action (\ref{eq:KKFermionLag}) and anticipated in (\ref{eq:A1A2A3KK}).  Further, some algebra allows one to verify that the KK tensors $A_3$ and $A_2$ in (\ref{eq:A3dddddd}) and (\ref{eq:A2KKduuu}) are indeed related as in (\ref{eq:A1A2A3KK}), in analogy with the $\cN=8$ relation (\ref{eq:A3A2}). 

The KK spin-$1/2$ fermion mass matrix  (\ref{eq:A3dddddd}) is infinite-dimensional and block-diagonal. The block at KK level $n$ is in the representations of SO(8) (or branchings thereof)
\begin{eqnarray} \label{eq:IrrepsA1KKSO8Spin1/2}
[1,0,1,0] \times [n,0,0,0]  & \longrightarrow & \underline{[n,0,0,1]}+  \underline{[n-1,0,1,0]} + [n+1,0,1,0] + [n-1,1,1,0] \nonumber \\
&& + [n-2,1,0,1]+ [n-2,0,0,1] \; , 
\end{eqnarray}
for solutions that uplift from the SO(8) gauging. For solutions that uplift from the ISO(7) gauging, the blocks lie in the following SO(7) representations (or their branchings):
\begin{eqnarray} \label{eq:IrrepsA1KKISO7Spin1/2}
([1,0,1] + [0,0,1])  \times [n,0,0]  & \longrightarrow & \underline{[n,0,1]}+  \underline{[n-1,0,1]} +[n+1,0,1] + [n-1,1,1] \hspace{30pt}  \\
&& + [n-2,1,1] +[n-2,0,1] + [n,0,1] + [n-1,0,1] \; ,\nonumber 
\end{eqnarray}
for $n=0, 1, 2, \ldots$  Again, only representations with non-negative Dynkin labels are actually present in both (\ref{eq:IrrepsA1KKSO8Spin1/2}) and (\ref{eq:IrrepsA1KKISO7Spin1/2}). Unlike its counterpart (\ref{eq:A1KKdd}) for the KK gravitini, the spin-$1/2$ KK fermion mass matrix (\ref{eq:A3dddddd}) contains unphysical states that must be removed from the spectrum. These correspond to the underlined representations in (\ref{eq:IrrepsA1KKSO8Spin1/2}) and (\ref{eq:IrrepsA1KKISO7Spin1/2}), which contain the Goldstini eaten by the massive gravitini at the same KK level $n$. Only the eigenvalues of (\ref{eq:A3dddddd}) that belong to representations not underlined in (\ref{eq:IrrepsA1KKSO8Spin1/2}) and (\ref{eq:IrrepsA1KKISO7Spin1/2}) constitute the physical KK fermion states of spin one-half.

%%%%%%%%%%%%%%%

\subsection{KK fermion shifts} \label{sec:KKShifts}

%%%%%%%%%%%%%%%

Like their $D=4$ $\cN=8$ counterparts $A_{1 i j}$, $A_{2i }{}^{jkl }$, the tensors $A_{1 \, i \Lambda , j \Sigma}$, $A_{2i \Lambda}{}^{jkl \Sigma}$ defined in (\ref{eq:A1KKdd}) and (\ref{eq:A2KKduuu})  are also `fermion shifts', in the sense that they analogously appear in the supersymmetry variations of the KK fermions. Indeed, the ExFT supersymmetry parameter can be expanded as
\begin{equation}\label{eq:SSsusy}
\bm{\epsilon}^i\left(x,Y\right) = \rho (Y )^{-\frac{1}{2}}  \,\epsilon^{i  \Lambda}(x)\, \mathcal{Y}_\Lambda (Y )  \; ,
\end{equation}
building again on \cite{Hohm:2014qga,Malek:2019eaz}. Inserting (\ref{eq:SSpertfermions}), (\ref{eq:SSsusy}) into the supersymmetry variations of the ExFT fermions \cite{Godazgar:2014nqa} a calculation analogous to \cite{Hohm:2014qga} using (\ref{eq:SU8connectionSS}) allows us to compute
\begin{equation} \label{eq:KKsusy}
\delta \psi_\mu^{i\Lambda} = 2 \, A_1{}^{i\Lambda , j \Sigma} \, \gamma_{\mu} \epsilon_{j \Sigma} + \ldots \; , \qquad
\delta\chi^{ijk \Lambda} = -2\sqrt{2} \,  A_{2 \, h \Sigma}{}^{ijk \Lambda} \, \epsilon^{ h \Sigma} + \ldots 
\end{equation}
with $A_1{}^{i\Lambda , j \Sigma}$ and $A_{2 \, h \Sigma}{}^{ijk \Lambda} $ respectively reproducing (\ref{eq:A1KKuu}) and (\ref{eq:A2KKduuu}). The terms shown here contain all the $\cN=8$ scalar contributions with no derivatives, while the ellipses hide contributions from the $\cN=8$ gauge fields, from derivatives of the scalars, and from derivatives of the supersymmetry parameter $\epsilon^{ i \Lambda}$. All the dependence on the internal coordinates drops from the ExFT supersymmetry variations when the coefficients of the harmonics ${\cal Y}_\Lambda$ is equated KK level by KK level on both sides of the equal sign, leaving the supersymmetry transformations  (\ref{eq:KKsusy}) for the KK fermions under which the full $D=4$ KK action is invariant.

%%%%%%%%%%%%%%%
%%%%%%%%%%%%%%%

\section{Complete KK spectra of $D=11$ $\cN=1$ AdS$_4$ solutions} \label{sec:N=1Spectrum11D}

%%%%%%%%%%%%%%%
%%%%%%%%%%%%%%%

We can now use our fermionic mass matrices to compute the KK fermion spectra of specific AdS$_4$ solutions. In this section, we will focus on AdS$_4$ solutions of $D=11$ supergravity \cite{Cremmer:1978km}  that uplift \cite{deWit:1986iy} from the $D=4$ $\cN=8$  SO(8) gauging \cite{deWit:1982ig}. For concreteness, we will restrict ourselves to the solutions that preserve at least the SU(3) subgroup of SO(8). These were classified in $D=4$ in \cite{Warner:1983vz} and uplifted to $D=11$ in \cite{Freund:1980xh,Corrado:2001nv,deWit:1984nz,Godazgar:2013nma,deWit:1984va,Englert:1982vs,Pope:1984bd}. The bosonic and fermionic KK level $n=0$  spectrum for these solutions is known (see \cite{Bobev:2010ib,Comsa:2019rcz}). Our results reproduce the fermionic spectra and extend them to higher KK levels $n \geq 1$. See table \ref{tab:KKGravitiniSO8} in appendix \ref{sec:SpecificSpectra} for a summary of the spectrum of KK gravitini for these solutions up to KK level $n=2$.

There are three supersymmetric solutions in this sector, with (super)symmetry $\cN=8$ SO(8) \cite{Freund:1980xh}, $\cN=2$ $\textrm{SU}(3) \times \textrm{U}(1)$ \cite{Warner:1983vz,Corrado:2001nv} and $\cN=1$ G$_2$ \cite{Warner:1983vz,deWit:1984nz,Godazgar:2013nma}. The complete supersymmetric KK spectrum for the former two is known \cite{Englert:1983rn,Sezgin:1983ik,Biran:1983iy,Klebanov:2008vq,Malek:2020yue,Varela:2020wty}, and we reproduce the corresponding fermionic sectors. For the $\cN=1$ G$_2$ solution, the  fermionic spectrum is new. Combining this with previously known sectors of the bosonic spectrum \cite{Pang:2017omp,Varela:2020wty}, we can determine the complete supersymmmetric spectrum for this $\cN=1$ solution.

\newpage

%%%%%%%%%%%%%%%

\subsection{Spectrum of the $\cN=1$ G$_2$-invariant solution } \label{sec:N=1G2SpectrumD=11}

%%%%%%%%%%%%%%%

The $\cN=1$ G$_2$-invariant AdS$_4$ solution was first found as a critical point of $D=4$ $\cN=8$ SO(8)-gauged supergravity \cite{deWit:1982ig} in \cite{Warner:1983vz}. The $S^7$ uplift of this solution to $D=11$ was determined in \cite{deWit:1984nz} and \cite{Godazgar:2013nma}. The first of these references provided the $D=11$ metric, while the second completed the uplift to include the three- and four-form fluxes. Geometrically, the $\cN=1$ G$_2$ solution in $D=11$ corresponds to a warped product of AdS$_4$ with a topological $S^7$. The metric on the latter can be written as a cohomogeneity-one, SO(7)-invariant deformation of the sine-cone metric foliated with round $S^6$ leaves. The latter is naturally equipped with its canonical, homogeneous nearly-K\"ahler structure. The three- and four-form fluxes can be written in terms of the nearly-K\"ahler forms, and break the SO(7) isometry down to a G$_2$ symmetry for the full solution. See \cite{Larios:2019kbw} for further details.

The KK spectrum of this solution is partially known. The spectrum at KK level $n=0$ may be determined by linearisation of $D=4$ $\cN=8$ SO(8) supergravity \cite{deWit:1982ig} around its $\cN=1$ G$_2$ critical point \cite{Warner:1983vz}. The $n=0$ scalar \cite{Bobev:2010ib}, spin-$1/2$ (see \cite{Comsa:2019rcz}), vector \cite{Borghese:2012qm} and gravitino (see \cite{Comsa:2019rcz}) spectra are thus known. The bosonic KK spectra at higher KK levels is also known partially. The KK graviton spectrum was calculated \cite{Dimmitt:2019qla} following the standard spin-2 methods of \cite{Bachas:2011xa}. The KK vector spectrum was computed \cite{Varela:2020wty} using ExFT techniques \cite{Malek:2019eaz,Malek:2020yue}. Now, we can use the fermionic mass matrices derived in section \ref{sec:KKFermionMassMat} to compute the spectrum of KK gravitini and spin-$1/2$ fermions. The results are summarised for the gravitini up to KK level $n=2$ in table \ref{tab:KKGravitiniSO8} of appendix \ref{sec:SpecificSpectra}. Further, we may use all these previous and new results to determine the complete $\cN=1$ KK spectrum about this AdS$_4$ solution, to which we now turn.

The complete KK spectrum arranges itself in representations of the residual supersymmetry and bosonic symmetry groups, $\textrm{OSp}(4|1) \times \textrm{G}_2$. These descend KK level by KK level from the OSp$(4|8)$ supermultiplets present in the spectrum at the $\cN=8$ SO(8) point \cite{Englert:1983rn} (see also {\it e.g.}~table 2 of \cite{Klebanov:2008vq} for a convenient summary). At fixed KK level $n$, the fields of all spins between 0 and 2 come in the (real) representations $[p,q]$ of G$_2$ determined by the branching $\textrm{G}_2 \subset \textrm{SO}(8)$. Fields in the same G$_2$ representations but different spin must then be allocated into OSp$(4|1)$ supermultiplets, starting from higher to lower spins. Table \ref{tab:OSp(4|1) supermultiplets} in the introduction comes in handy to carry out this exercise. Only the MGRAV and MVEC (please refer to the table for the acronyms used throughout) OSp$(4|1)$ supermultiplets  have their conformal dimensions $E_0$ fixed in terms of the spin $s_0$ of the superconformal primary. For all other multiplets present in the spectrum, $E_0$ cannot be determined from group theory alone. Satisfactorily enough, the dimensions and their multiplicities computed from the previously known bosonic \cite{Dimmitt:2019qla,Varela:2020wty} and from our new fermionic mass spectra match this $\textrm{OSp}(4|1) \times \textrm{G}_2$ structure and bring in the dimensions $E_0$.

We find that the complete supersymmetric KK spectrum of the $\cN=1$ G$_2$ invariant solution \cite{Warner:1983vz,deWit:1984nz,Godazgar:2013nma} has the following structure. At KK level $n=0$, there are, as expected, a MGRAV and a MVEC, which respectively lie in the trivial and the adjoint representations of G$_2$. The former corresponds to the $\cN=1$ supergravity multiplet, which includes the massless graviton and gravitino. The latter contains the vectors that gauge the residual symmetry G$_2$, along with their spin-$1/2$ superpartners. KK level $n=0$ is completed with a GINO multiplet containing the $\mathbf{7}$ massive gravitini, along with $\bm{1}$ and $\bm{27}$ CHIRAL multiplets. Higher KK levels $n\geq 1$ contain all four generic supermultiplets, GRAV, GINO, VEC and CHIRAL, in suitable representations of G$_2$. The supersymmetric KK spectrum for the first four levels, $n=0,1,2,3$, is summarised in tables \ref{tab:multipletsatlevel0G2D11}--\ref{tab:multipletsatlevel3G2D11} below. For each OSp$(4|1)$ supermultiplet with given G$_2$ quantum numbers $[p,q]$, the dimension $E_0$ is shown next to the corresponding acronym. An entry of the form $m \times (E_0)$ indicates that there are $m$ such supermultiplets. Whenever there is only one multiplet, $m=1$, we simply write $(E_0)$.

\vspace{30pt}

\begin{table}[H]
\begin{center}
{\footnotesize
\begin{tabular}{|p{30mm}|p{30mm}|} 					\hline
$[0,0]$ 				& 	$[0,1]$ 							\\[1pt]
MGRAV $\left( \tfrac52 \right)$ 			& MVEC $\left( \tfrac32 \right)$  	\\
 CHIRAL  $\left( 1+\sqrt{6} \right)$	& 		 				\\[5pt] \hline
$[1,0]$ 				 			\\[1pt]
GINO $\left( 1+\tfrac{\sqrt{6}}{2} \right) $ 				\\[5pt] \cline{1-1}
$[2,0]$ 			\\[1pt]
CHIRAL $\left(  1+\tfrac{\sqrt{6}}{6}  \right)$ 	\\[5pt] \cline{1-1}
\end{tabular}
}
\caption{\footnotesize{Supermultiplets at KK level $n=0$ for the $D=11$ $\cN=1$ G$_2$-invariant AdS$_4$ solution.}\normalsize}
\label{tab:multipletsatlevel0G2D11}
\end{center}
\end{table}

\vspace{30pt}

\begin{table}[H]
\begin{center}
{\footnotesize
\begin{tabular}{|p{35mm}|p{30mm}|} 					\hline
$[0,0]$ 							& 	$[0,1]$ 							\\[1pt]
GRAV $\left( 1+\frac{\sqrt{106}}{4} \right)$ 				& GINO $\left( 1+\frac{3}{4}\sqrt{6} \right)$  	\\
 CHIRAL$\left( 1+\frac{\sqrt{166}}{4} \right)$ 								& VEC $ \left( 1+\frac{\sqrt{74}}{4} \right) $		 				\\[5pt] \hline
$[1,0]$ 							& 	$[1,1]$ 							\\[1pt]
GRAV $\left( 1+\tfrac{\sqrt{66}}{4} \right)$ 			& VEC $\left( 1+ \tfrac{\sqrt{14}}{4} \right)$  	\\
VEC $\left( 1+ \tfrac{\sqrt{114}}{4}  \right)$ 		& 		 		\\
CHIRAL $\left( 1+ \tfrac{3}{2}\sqrt{\tfrac{7}{2}}  \right)$ 		& 		 				\\[1pt]
GINO $\left( 1+ \tfrac{\sqrt{94}}{4}  \right)$  &    \\[5pt] \hline
$[2,0]$ 				 			\\[1pt]
GINO $\left( 1+\tfrac{1}{2} \sqrt{\tfrac{61}{6}} \right) $ \\[1pt]
VEC $\left( 1+ \sqrt{\frac{91}{24}}\right) $ 				\\
CHIRAL  $\left( 1+ \tfrac{\sqrt{654}}{12} \right) $ \\[5pt] \cline{1-1}
$[3,0]$ 			\\[1pt]
CHIRAL $\left( 1+\sqrt{\frac{3}{8}} \right)$ 	\\[5pt] \cline{1-1}
\end{tabular}
}
\caption{\footnotesize{Supermultiplets at KK level $n=1$ for the $D=11$ $\cN=1$ G$_2$-invariant AdS$_4$ solution.}\normalsize}
\label{tab:multipletsatlevel1G2D11}
\end{center}
\end{table}

\begin{table}[H]
\begin{center}
{\footnotesize
\begin{tabular}{|p{35mm}|p{35mm}|p{35mm}|} 					\hline
$[0,0]$ 										& 	$[0,1]$ 									& 	$[0,2]$ 			\\[1pt]
GRAV $\left(  \frac{9}{2} \right)  $ 			& GINO $2\times\left(  4 \right) $ 			& CHIRAL $\left(  1+ \sqrt{\tfrac{7}{2}} \right) $ 	\\
 CHIRAL  $2\times\left(  5 \right)  $ 						&VEC $ 2\times\left(  1+ \sqrt{\tfrac{41}{4}} \right) $ & 							\\[5pt] \hline
$[1,0]$ 									& 	$[1,1]$ 			\\[1pt]
GRAV $\left(  1+ \tfrac{\sqrt{39}}{2} \right) $ 		& GINO $\left( 1+ \tfrac{\sqrt{21}}{2} \right)$ 			\\ 
GINO $2\times\left( 1+\sqrt{\frac{23}{2}} \right) $ 		& VEC $2\times\left(  1+ \sqrt{\frac{13}{2}} \right)$ 	\\
VEC $2\times\left( 1+\sqrt{\tfrac{51}{4}} \right) $ 		&  CHIRAL $\left(  1+ \tfrac{\sqrt{29}}{2} \right)$ \\
CHIRAL $\left( 1+\sqrt{\tfrac{27}{2}} \right) $ 		&   \\[5pt]  \cline{1-2}
$[2,0]$ 									& 	$[2,1]$ 			\\[1pt]
GRAV $\left( 1+ \frac{\sqrt{231}}{6} \right)  $ 		& VEC $\left( 1+ \sqrt{\tfrac{23}{12}} \right)$ 			\\[5pt] 
GINO $\left(  1+ \tfrac{7}{\sqrt{6}} \right)  $ 		& 		\\[5pt] 
VEC $2\times \left(  1+ \sqrt{\tfrac{113}{12}} \right)  $ & 	\\
CHIRAL $ 3\times\left(  1+ \sqrt{\tfrac{61}{6}} \right)  $ & 	\\[5pt] \cline{1-2}
$[3,0]$ 			\\[1pt]
GINO $\left(  3 \right) $\\
VEC $ \left(  1+ \sqrt{\tfrac{21}{4}} \right) $ \\
CHIRAL $\left(  1+ \sqrt{6} \right) $	\\[5pt] \cline{1-1}
$[4,0]$ 			\\[1pt]
CHIRAL $\left( 2 \right) $ 	\\[5pt] \cline{1-1}
\end{tabular}
}
\caption{\footnotesize{Supermultiplets at KK level $n=2$ for the $D=11$ $\cN=1$ G$_2$-invariant AdS$_4$ solution.}\normalsize}
\label{tab:multipletsatlevel2G2D11}
\end{center}
\end{table}

\begin{table}[H]
\begin{center}
{\footnotesize
\begin{tabular}{|p{40mm}|p{35mm}|p{35mm}|} 					\hline
$[0,0]$ 										& 	$[0,1]$ 									& 	$[0,2]$ 			\\[1pt]
GRAV $\left(  1+\frac{3}{4}\sqrt{34} \right)  $ 			&GINO $2\times\left(  1+ \frac{\sqrt{254}}{4} \right) $ 			& VEC $\left(  1+ \frac{\sqrt{154}}{4}\right) $ 	\\
 CHIRAL $2\times\left(  1+\frac{\sqrt{366}}{4} \right)  $ 						& VEC $3\times\left(  1+\frac{\sqrt{274}}{4} \right) $ & 		CHIRAL	$\left(  1+ \frac{\sqrt{166}}{4}\right) $				\\[5pt] \hline
$[1,0]$ 									& 	$[1,1]$ 	& $[1,2]$		\\[1pt]
GRAV $\left(  1+ \tfrac{\sqrt{266}}{4} \right) $ 		& GINO $2\times\left( 1+ \tfrac{\sqrt{194}}{4} \right)$ 	& 	CHIRAL	$\left(  1+ \frac{\sqrt{86}}{4}\right) $				\\ 
GINO $2\times\left( 1+\frac{7\sqrt{6}}{4} \right) $ 		& VEC $3\times\left(  1+ \frac{\sqrt{214}}{4} \right)$ &	\\
VEC $2\times\left( 1+\frac{\sqrt{314}}{4} \right) $ 		&CHIRAL $2\times\left(  1+ \tfrac{\sqrt{226}}{4} \right)$ & \\[1pt]
CHIRAL $2\times\left(  1+ \tfrac{\sqrt{326}}{4} \right)$ 		& &   \\[5pt]  \cline{1-3}
$[2,0]$ 									& 	$[2,1]$ 			\\[1pt]
GRAV $\left( 1+ \sqrt{\frac{319}{24}} \right)  $ 		& GINO $\left( 1+ \sqrt{\tfrac{181}{24}} \right)$ 			\\[5pt] 
GINO $2\times\left(  1+ \tfrac{19}{2\sqrt{6}} \right)  $ 		& 	 VEC $2\times\left(  1+ \sqrt{\frac{211}{24}} \right)$	\\[5pt] 
VEC $3\times \left(  1+ \sqrt{\tfrac{391}{24}} \right)  $ & 	CHIRAL $ \left(  1+ \sqrt{\tfrac{229}{24}} \right)  $\\
CHIRAL $ 3\times\left(  1+ \sqrt{\tfrac{409}{24}} \right)  $ & 	\\[5pt] \cline{1-2}
$[3,0]$ 	& $[3,1]$		\\[1pt]
GRAV $ \left(  1+ \tfrac{\sqrt{146}}{4} \right) $ &      VEC $ \left(  1+ \tfrac{3\sqrt{6}}{4} \right) $              \\
GINO           $ \left(  1+ \tfrac{\sqrt{174}}{4} \right) $ & \\
VEC $ 2\times\left(  1+ \tfrac{\sqrt{194}}{4} \right) $ & \\
CHIRAL $3\times\left(  1+ \frac{\sqrt{206}}{4}\right) $ &	\\[5pt] \cline{1-2}
$[4,0]$ 			\\[1pt]
GINO     $ \left(  1+ \tfrac{\sqrt{94}}{4} \right) $                 \\
VEC $ \left(  1+ \tfrac{\sqrt{114}}{4} \right) $                  \\
CHIRAL $\left( 1+\frac{3}{2}\sqrt{\frac{7}{2}} \right) $ 	\\[5pt] \cline{1-1}
$[5,0]$       \\[1pt]
CHIRAL $ \left(  1+ \tfrac{7}{2\sqrt{6}} \right) $ \\[5pt] \cline{1-1}
\end{tabular}
}
\caption{\footnotesize{Supermultiplets at KK level $n=3$ for the $D=11$ $\cN=1$ G$_2$-invariant AdS$_4$ solution.}\normalsize}
\label{tab:multipletsatlevel3G2D11}
\end{center}
\end{table}

 \begin{table}[H]

\centering

\scriptsize{

\begin{tabular}{llccclcl}
\hline
\multicolumn{2}{c}{$\Delta$}  && $n$  &&   $\textrm{G}_2  $  && $\cN=1$ supermultiplet
            \\ \hline
 %
 %
 %
 %
 % 
%
% \hline
 %
 %
$ 1 + \sqrt{\tfrac{3}{8}} $ & $1.38$ & &  1   &&  $[3,0]     $   && CHIRAL $\left( 1 + \sqrt{\frac{3}{8}}   \right)  $  \\[5pt] 
 %
%
% \hline
 %
 %
$ 1 +\tfrac{\sqrt{6}}{6}  $ & $1.41$ & &  0   &&  $[2,0]     $   && CHIRAL $\left( 1 + \tfrac{\sqrt{6}}{6}   \right)  $  \\[5pt] 
%
% \hline
%
%
$ 2 $ & $2.00$ & & 2  &&  $[4,0]    $   && CHIRAL $\left( 2  \right)  $  \\[5pt] 
 %
%
% \hline
 %
$ 2 + \sqrt{\tfrac{3}{8}} $ & $2.38$ & &  1  &&  $[3,0]     $   && CHIRAL $\left( 1 + \sqrt{\tfrac{3}{8}}    \right)  $  \\[5pt] 
 %
%
%
 % \hline
 %
 %
$ 2 +\tfrac{\sqrt{6}}{6}  $ & $2.41$ & &  0   &&  $[2,0]     $   && CHIRAL $\left( 1 + \tfrac{\sqrt{6}}{6}   \right)  $  \\[5pt] 
$ 1 + \tfrac{7}{2\sqrt{6}} $ & $2.43$ & & 3   &&  $[5,0]  $   && CHIRAL $\left( 1 + \tfrac{7}{2\sqrt{6}} \right) $ \\[5pt] 
 %
%
 % \hline
 %
 %
$ \tfrac32 + \tfrac{\sqrt{14}}{4} $ & $2.44$ & & 1   &&  $[1,1]  $   && VEC $\left( 1 +  \tfrac{\sqrt{14}}{4} \right) $ \\[5pt] 
 %
%
%
 % \hline
 %
 %
$ 1 +\sqrt{ \tfrac{7}{2}} $ & $2.87$ & & 2   &&  $[0,2]  $   && CHIRAL $\left( 1 +\sqrt{ \tfrac{7}{2}}  \right) $ \\[5pt] 
$ \tfrac32 + \sqrt{\frac{23}{12}} $ & $2.88$ & & 2   &&  $[2,1]  $   && VEC $\left( 1 + \sqrt{\frac{23}{12}}  \right) $ \\[5pt] 
$3$ & $3.00$ & &  2   &&  $[4,0]  $   && CHIRAL $\left( 2 \right)   $   \\[5pt] 
 \hline
\end{tabular}

}\normalsize

\caption{\scriptsize{All KK scalars with dimension $\Delta \leq 3$ around the $D=11$ $\cN=1$ G$_2$-invariant AdS$_4$ solution.  }\normalsize}
\label{tab:N=1G2SO(8)calars}
\end{table}

From tables \ref{tab:multipletsatlevel0G2D11}--\ref{tab:multipletsatlevel3G2D11} it is possible to infer that the conformal dimension $E_0$ for each of the OSp$(4|1)$ supermultiplets present in the spectrum at KK level $n=0 , 1 , 2 \ldots $, with G$_2$ quantum numbers $[p,q]$, is
{\setlength\arraycolsep{2pt}
\begin{eqnarray} \label{eq:SpecGRAVG2D11}
 \textrm{(M)GRAV}  & : & E_0 = 1 + \sqrt{ \tfrac94 + \tfrac{5}{8} n (n+6)-\tfrac54 \, {\cal C}_2 ( p,q )  }  ,   \hspace{10pt} \\[5pt]
\label{eq:SpecGINOG2D11}
 \textrm{GINO}  & : & E_0 = 1 + \sqrt{ 4 + \tfrac58 n (n+6) - \tfrac54 \, {\cal C}_2 ( p,q ) } \; ,   \hspace{10pt} \\[5pt]
\label{eq:SpecVECG2D11}
 \textrm{(M)VEC}  & : & E_0 = 1 + \sqrt{ \tfrac{21}{4} + \tfrac58 n (n+6) - \tfrac54 \, {\cal C}_2 ( p,q ) } \; ,   \hspace{10pt} \\[5pt]
\label{eq:SpecCHIRALG2D11}
 \textrm{CHIRAL} & : & E_0 = 1 + \sqrt{ 6 + \tfrac58 n (n+6) - \tfrac54 \, {\cal C}_2 ( p,q ) } \; .
\end{eqnarray}
In these expressions, $n(n+6)$ are the eigenvalues of the scalar Laplacian on $S^7$, and\footnote{ \label{fn:NormCas}
Recall that the overall normalisation of the Casimir operator is arbitrary.  A popular normalisation, which we use for the eigenvalue ${\cal C}_2$ of the G$_2$ (\ref{eq:G2CasimirEigenv}), SU(3) (\ref{eq:SU3CasimirEigenv}), SO(7) (\ref{eq:SO(7)Casimir}) and SO(6) (\ref{eq:SU(4)Casimir}) quadratic Casimir operator in the representation $R$, is ${\cal C}_2 = d_G/( 2 \, d_R ) \, {\cal I}_R$, where $d_R$, $d_G$ and ${\cal I}_R$ are respectively the dimension of $R$, the dimension of the adjoint, and the Dynkin index of $R$, see {\it e.g.} \cite{Slansky:1981yr}.
} 
\begin{equation} \label{eq:G2CasimirEigenv}
{\cal C}_2( p,q ) \equiv \tfrac13 \, p(p+5) +  q(q+3) +  \, pq \; , 
\end{equation}
is the eigenvalue of the quadratic Casimir operator of G$_2$ in the $[p,q]$ representation. The dimension (\ref{eq:SpecGRAVG2D11}) of the (M)GRAV supermultiplets is in agreement with the masses of the individual graviton states given in table 2 of \cite{Dimmitt:2019qla}, with $n_\textrm{here} = p_\textrm{here} = k_\textrm{there}$, $q_\textrm{here} = 0$. Likewise, the individual vector states contained in the supermultiplets with dimensions (\ref{eq:SpecGRAVG2D11})--(\ref{eq:SpecVECG2D11}) match the masses reported in table 14 of \cite{Varela:2020wty} up to second KK level. More generally, recalling  from table \ref{tab:OSp(4|1) supermultiplets} the value of the conformal primary spin $s_0$ for each supermultiplet, the formulae (\ref{eq:SpecGRAVG2D11})--(\ref{eq:SpecCHIRALG2D11}) can be collectively written as in equation (\ref{eq:E0generic}) with $d=7$, $\alpha = \tfrac58$ and $\beta = \tfrac54$ therein, as already advertised in the introduction. 

\newpage 

The spectrum of individual KK scalar states can be inferred from the complete supersymmetric KK spectrum. Table \ref{tab:N=1G2SO(8)calars} lists  all the scalars with conformal  dimensions $\Delta \leq 3$. The table includes the analytical value of $\Delta$ together with a convenient numerical approximation. Also shown in the table is the KK level $n$ at which each scalar appears, as well as its $\textrm{G}_2$ charges $[p,q]$. The OSp$(4|1)$ supermultiplet with dimension $E_0$, at the same KK level $n$ and with the same G$_2$ charges $[p,q]$, that contains each scalar is also shown. The dimension $\Delta$ will only match $E_0$ if the scalar in question is the superconformal primary of its multiplet. The scalars in table \ref{tab:N=1G2SO(8)calars} are dual to relevant ($\Delta < 3$) or classically marginal ($\Delta = 3$) operators in the dual field theory. All scalars with $\Delta \leq 3$ turn out to arise at KK levels $n=0,1,2,3$. Each of these KK levels contain scalars dual to irrelevant ($\Delta >3$) operators as well. At KK levels $n \geq 4$, all scalars are dual to irrelevant operators.

%%%%%%%%%%%%%%%
%%%%%%%%%%%%%%%

\section{Complete KK spectra of $\cN=1$ AdS$_4$ solutions of type IIA} \label{sec:N=1Spectrum10D}

%%%%%%%%%%%%%%%
%%%%%%%%%%%%%%%

Next we turn to compute the KK fermionic spectra of the AdS$_4$ solutions of massive type IIA supergravity  \cite{Romans:1985tz} that are obtained from vacua of $D=4$ $\cN=8$ dyonic ISO(7) supergravity \cite{Guarino:2015qaa} upon consistent uplift on $S^6$ \cite{Guarino:2015jca,Guarino:2015vca}. Again for definiteness, we will focus on those solutions that preserve at least the SU(3) subgroup of SO(7). From a four-dimensional perspective, these were classified in \cite{Guarino:2015qaa}. Their type IIA uplifts were given in \cite{Guarino:2015jca,Guarino:2015vca,Varela:2015uca}. Using our mass matrices from section \ref{sec:KKFermionMassMat}, we have computed the spectrum of KK gravitini and spin-$1/2$ fermions for the first few KK levels. A summary of the gravitino spectra of these solutions up to KK level $n=2$ is provided in table \ref{tab:KKGravitiniISO7} of appendix \ref{sec:SpecificSpectra}. 

This sector contains three supersymmetric solutions with residual (super)symmetry $\cN=2$ $\textrm{SU}(3) \times \textrm{U}(1)$ \cite{Guarino:2015jca}, $\cN=1$ G$_2$ \cite{Borghese:2012qm,Behrndt:2004km,Guarino:2015vca}, and $\cN=1$ SU(3) \cite{Borghese:2012zs,Guarino:2015qaa,Varela:2015uca}. The complete supersymmetric spectrum of the $\cN=2$ solution was recently obtained in \cite{Varela:2020wty}, and our results match the fermionic spectrum that can be inferred from the results of that reference. Here, we will give the complete supersymmetric spectrum of the $\cN=1$ solutions. We now move on to discuss them in turn.

%%%%%%%%%%%%%%%

\subsection{Spectrum of the $\cN=1$ G$_2$-invariant solution } \label{sec:N=1G2SpectrumIIA}

%%%%%%%%%%%%%%%

The AdS$_4$ solution with $\cN=1$ supersymmetry and G$_2$ symmetry was first reported as a critical point of $D=4$ $\cN=8$ dyonic ISO(7) supergravity in \cite{Borghese:2012qm}. The $S^6$ uplift to massive IIA \cite{Romans:1985tz} was performed in \cite{Guarino:2015vca,Varela:2015uca} (see (4.6) of the latter reference), and shown to coincide with a previously known solution first written in \cite{Behrndt:2004km}. The ten-dimensional solution is a direct product of AdS$_4$ and the round $S^6$ sphere, endowed with its SO(7)-invariant homogeneous Einstein metric. The latter is inherited from the canonical homogeneous nearly-K\"ahler structure on $S^6$. All type IIA forms are active and take values along the nearly-K\"ahler forms, thereby reducing the symmetry of the full solution to G$_2$.  

Some details of the KK spectrum about this solution are already known. As always, the spectrum at KK level $n=0$ is found by linearising $D=4$ $\cN=8$ ISO(7) supergravity around the $\cN=1$ G$_2$ critical  point \cite{Borghese:2012qm}. The $n=0$ scalar and vector spectrum was given in that reference. The $n=0$ fermion spectrum, which can be deduced by supersymmetry from the $n=0$ bosonic spectrum \cite{Borghese:2012qm}, has been explicitly given in the recent \cite{Bobev:2020qev}. At higher KK levels $n \geq 1$, the spectrum of KK gravitons \cite{Pang:2017omp} and vectors \cite{Varela:2020wty} is also known. These were respectively computed using standard spin-2 techniques \cite{Bachas:2011xa} and ExFT methods \cite{Malek:2019eaz,Malek:2020yue}. Here, we have computed the spectrum of KK gravitini and spin-$1/2$ fermions using the mass matrices (\ref{eq:A1KKdd}) and (\ref{eq:A3dddddd}). Combining these new results with the previously known ones, we can further obtain the complete $\cN=1$ KK spectrum of this solution.
 
Like in the $D=11$ case of section \ref{sec:N=1G2SpectrumD=11}, the complete KK spectrum combines itself in representations of $\textrm{OSp}(4|1) \times \textrm{G}_2$. The process to find the multiplet structure of the spectrum is very similar to that explained in detail in section \ref{sec:N=1G2SpectrumD=11} and, for that reason, we shall be brief. The most important difference with respect to the $D=11$ case is that the G$_2$ representations at fixed KK level $n$ branch from the putative SO(7) representations summarised in table 1 of \cite{Varela:2020wty}. Proceeding as explained in section  \ref{sec:N=1G2SpectrumD=11}, we find the $\textrm{OSp}(4|1) \times \textrm{G}_2$ structure of the spectrum. Except for the massless multiplets, the dimensions $E_0$ of the multiplets are again left undetermined by group theory. Fortunately, the dimensions and their multiplicities of the known bosonic KK fields \cite{Pang:2017omp,Varela:2020wty} and the new fermionic KK fields are compatible with this $\textrm{OSp}(4|1) \times \textrm{G}_2$ structure and allow us to give explicitly the supermultiplet dimensions.

At KK level $n=0$, the spectrum of the G$_2$ $\cN=1$ solution \cite{Borghese:2012qm,Behrndt:2004km,Guarino:2015vca} of type IIA coincides with the spectrum of its $D=11$ counterpart \cite{Warner:1983vz,deWit:1984nz,Godazgar:2013nma}, as already noted in \cite{Borghese:2012qm}. This spectrum is summarised in table \ref{tab:multipletsatlevel0G2IIA} below, which is included for completeness although this table is identical to table \ref{tab:multipletsatlevel0G2D11} for the $D=11$ case. The spectra in the IIA and $D=11$ cases do differ at higher KK levels: see tables \ref{tab:multipletsatlevel1G2IIA}--\ref{tab:multipletsatlevel3G2IIA} below for the supersymmetric spectrum at levels $n=1,2,3$ in the type IIA case. The dimension $(E_0)$ is shown next to each supermultiplet. An entry of the form $m \times (E_0)$ indicates that there are $m$ such supermultiplets, with the label $m$ omitted when $m=1$.

\vspace{30pt}

%Multiplets at level 0

\begin{table}[H]
\begin{center}
{\footnotesize
\begin{tabular}{|p{30mm}|p{30mm}|} 					\hline
$[0,0]$ 				& 	$[0,1]$ 							\\[1pt]
MGRAV $\left( \tfrac52 \right)$ 			& MVEC $\left( \tfrac32 \right)$  	\\
CHIRAL  $\left( 1+\sqrt{6} \right)$ 		& 		 				\\[5pt] \hline
$[1,0]$ 				 			\\[1pt]
GINO $\left( 1+\tfrac{\sqrt{6}}{2} \right) $ 				\\[5pt] \cline{1-1}
$[2,0]$ 			\\[1pt]
CHIRAL $\left(  1+\tfrac{\sqrt{6}}{6}  \right)$ 	\\[5pt] \cline{1-1}
\end{tabular}
}
\caption{\footnotesize{Supermultiplets at KK level $n=0$ for the $\cN=1$ G$_2$-invariant AdS$_4$ solution of type IIA.}\normalsize}
\label{tab:multipletsatlevel0G2IIA}
\end{center}
\end{table}

\begin{table}[H]
\begin{center}
{\footnotesize
\begin{tabular}{|p{30mm}|p{30mm}|} 					\hline
$[0,0]$ 							& 	$[0,1]$ 							\\[1pt]
GINO $\left( 4 \right)$ 				& GINO $ \left( 3 \right) $ 	\\
 								& CHIRAL $\left( 1+\sqrt{6} \right)$ 		 				\\[5pt] \hline
$[1,0]$ 							& 	$[1,1]$ 							\\[1pt]
GRAV $\left( 1+\tfrac{\sqrt{19}}{2} \right)$ 			& VEC $\left( 1+ \sqrt{ \tfrac32 } \right)$  	\\
VEC $\left( 1+ \tfrac{\sqrt{31}}{2}  \right)$ 		& 		 		\\
CHIRAL $\left( 1+ \tfrac{\sqrt{34}}{2}  \right)$ 		& 		 				\\[5pt] \hline
$[2,0]$ 				 			\\[1pt]
GINO $\left( 1+ \sqrt{\tfrac{19}{6}} \right) $ \\[1pt]
VEC $\left( 1+ \tfrac{\sqrt{159}}{6} \right) $ 				\\[5pt] \cline{1-1}
$[3,0]$ 			\\[1pt]
CHIRAL $\left( 2 \right)$ 	\\[5pt] \cline{1-1}
\end{tabular}
}
\caption{\footnotesize{Supermultiplets at KK level $n=1$ for the $\cN=1$ G$_2$-invariant AdS$_4$ solution of type IIA.}\normalsize}
\label{tab:multipletsatlevel1G2IIA}
\end{center}
\end{table}

\vspace{30pt}

\begin{table}[H]
\begin{center}
{\footnotesize
\begin{tabular}{|p{35mm}|p{35mm}|p{35mm}|} 					\hline
$[0,0]$ 										& 	$[0,1]$ 									& 	$[0,2]$ 			\\[1pt]
CHIRAL $\left(  1+ \sqrt{\tfrac{53}{3}} \right)$ 			& GINO $\left(  1+ \tfrac{4\sqrt{6}}{3} \right) $ 			& CHIRAL $\left(  1+ \tfrac{\sqrt{186}}{6} \right) $ 	\\
 											& VEC $\left(  1+ \tfrac{\sqrt{429}}{6} \right) $ & 							\\[5pt] \hline
$[1,0]$ 									& 	$[1,1]$ 			\\[1pt]
GINO $\left(  1+ \tfrac{\sqrt{474}}{6} \right) $ 		& GINO $\left( 1+ \tfrac{\sqrt{249}}{6} \right)$ 			\\[5pt] 
VEC $\left( 1+ \tfrac{\sqrt{519}}{6} \right) $ 		& VEC $\left(  1+ \tfrac{7\sqrt{6}}{6} \right)$ 	\\[5pt] 
								 		& CHIRAL $\left(  1+ \tfrac{\sqrt{321}}{6} \right)$ 			\\[5pt] \cline{1-2}
$[2,0]$ 									& 	$[2,1]$ 			\\[1pt]
GRAV $\left(  1+ \tfrac{\sqrt{291}}{6} \right)  $ 		& VEC $\left( 1+ \tfrac{\sqrt{129}}{6} \right)$ 			\\[5pt] 
VEC $\left(  1+ \tfrac{\sqrt{399}}{6} \right)  $ 		& 		\\[5pt] 
CHIRAL $2 \times \left(  1+ \sqrt{\tfrac{71}{6}} \right)  $ & 	\\[5pt] \cline{1-2}
$[3,0]$ 			\\[1pt]
GINO $ \left(  1+ \sqrt{\tfrac{17}{3}} \right) $ \\
VEC $\left(  1+ \tfrac{\sqrt{249}}{6} \right) $	\\[5pt] \cline{1-1}
$[4,0]$ 			\\[1pt]
CHIRAL $\left(  1+ \tfrac{2\sqrt{6}}{3} \right) $ 	\\[5pt] \cline{1-1}
\end{tabular}
}
\caption{\footnotesize{Supermultiplets at KK level $n=2$ for the $\cN=1$ G$_2$-invariant AdS$_4$ solution of type IIA.}\normalsize}
\label{tab:multipletsatlevel2G2IIA}
\end{center}
\end{table}

\begin{table}[H]
\begin{center}
{\footnotesize
\begin{tabular}{p{35mm}|p{35mm}|p{35mm}|}

			\cline{2-3} %\hline 
 											& 	$[0,1]$ 									& 	$[0,2]$ 			\\[1pt]
 											& VEC $\left(  \tfrac{11}{2} \right) $ 					& VEC $\left(  1+ \tfrac{\sqrt{51}}{2} \right) $ 	\\
 											&  & 							\\[5pt] \hline
\multicolumn{1}{ |l|  }{$[1,0]$	}									& 	$[1,1]$ 									& 	$[1,2]$		\\[1pt]
\multicolumn{1}{ |l| }{CHIRAL $\left(  1+ \sqrt{\tfrac{47}{2}} \right)$} 			& GINO $\left(  1+ \tfrac{\sqrt{61}}{2} \right) $ 			& CHIRAL $\left(  1+ \tfrac{\sqrt{34}}{2} \right) $ 	\\
\multicolumn{1}{ |l| }{}										& VEC $\left(  1+ \sqrt{\tfrac{33}{2}} \right) $ & 				\\
\multicolumn{1}{ |l| }{}										&	 CHIRAL $\left(  1+ \tfrac{\sqrt{69}}{2} \right) $ & 							\\[5pt] \hline
\multicolumn{1}{ |l|  }{$[2,0]$} 									& 	$[2,1]$ 			\\[1pt]
\multicolumn{1}{ |l| }{GINO $\left(  1+ \sqrt{\tfrac{109}{6}} \right) $} 		& GINO $\left(   1+ \tfrac{4\sqrt{6}}{3}   \right)$ 			\\[5pt] 
\multicolumn{1}{ |l|  }{VEC $\left( 1+ \tfrac{\sqrt{699}}{6} \right) $} 		& VEC $\left(  1+ \tfrac{\sqrt{429}}{6} \right)$ 	\\[5pt] 
\multicolumn{1}{ |l|  }{}								 		& CHIRAL $\left(  1+ \tfrac{\sqrt{114}}{3} \right)$ 			\\[5pt] \cline{1-2}
\multicolumn{1}{ |l|  }{$[3,0]$} 									& 	$[3,1]$ 			\\[1pt]
\multicolumn{1}{ |l| }{GRAV $\left( \tfrac92 \right)  $} 					& VEC $\left( 1+ \tfrac{\sqrt{26}}{2} \right)$ 			\\[5pt] 
\multicolumn{1}{ |l|  }{VEC $\left(  1+ \tfrac{\sqrt{61}}{2} \right)  $} 		& 		\\[5pt] 
\multicolumn{1}{ |l| }{CHIRAL $2 \times \left(  5 \right)  $} & 	\\[5pt] \cline{1-2}
\multicolumn{1}{ |l|  }{$[4,0]$} 			\\[1pt]
\multicolumn{1}{ |l|  }{GINO $ \left(  4  \right) $} \\
\multicolumn{1}{ |l|  }{VEC $\left(  1+ \tfrac{\sqrt{41}}{2} \right) $}	\\[5pt] \cline{1-1}
\multicolumn{1}{ |l|  }{$[5,0]$} 			\\[1pt]
\multicolumn{1}{ |l|  }{CHIRAL $\left(  1+ \sqrt{\tfrac{31}{6}}  \right) $ }	\\[5pt] \cline{1-1}
\end{tabular}
}
\caption{\footnotesize{Supermultiplets at KK level $n=3$ for the $\cN=1$ G$_2$-invariant AdS$_4$ solution of type IIA.}\normalsize}
\label{tab:multipletsatlevel3G2IIA}
\end{center}
\end{table}

Inspection of tables \ref{tab:multipletsatlevel0G2IIA}--\ref{tab:multipletsatlevel3G2IIA} allows us to deduce generic expressions for the conformal dimensions of the OSp$(4|1)$ supermultiplets contained in the KK spectrum. Each type of supermultiplet in the $[p,q]$ representation of G$_2$ at KK level $n=0,1,2, \ldots$ has the following scaling dimension:
{\setlength\arraycolsep{2pt}
\begin{eqnarray} \label{eq:SpecGRAVG2IIA}
 \textrm{(M)GRAV}  & : & E_0 = 1 + \sqrt{ \tfrac94 + \tfrac{5}{12} n (n+5) }  ,   \hspace{10pt} \\[5pt]
\label{eq:SpecGINOG2IIA}
 \textrm{GINO}  & : & E_0 = 1 + \sqrt{ 4 + \tfrac56 n (n+5) - \tfrac54 \, {\cal C}_2 (p,q) } \; ,   \hspace{10pt} \\[5pt]
\label{eq:SpecVECG2IIA}
 \textrm{(M)VEC}  & : & E_0 = 1 + \sqrt{ \tfrac{21}{4} + \tfrac56 n (n+5) - \tfrac54 \, {\cal C}_2 (p,q) } \; ,   \hspace{10pt} \\[5pt]
\label{eq:SpecCHIRALG2IIA}
 \textrm{CHIRAL} & : & E_0 = 1 + \sqrt{ 6 + \tfrac56 n (n+5) - \tfrac54 \, {\cal C}_2 (p,q) } \; .
\end{eqnarray}
}Here, $n(n+5)$ are the eigenvalues of the scalar Laplacian on $S^6$ and ${\cal C}_2( p,q )$ are the eigenvalues (\ref{eq:G2CasimirEigenv}) of the quadratic Casimir operator of G$_2$ in the $[p,q]$ representation. The (M)GRAV supermultiplets have dimensions (\ref{eq:SpecGRAVG2IIA}) that agree with the individual graviton masses given in (3.1) of \cite{Pang:2017omp} with $n_\textrm{here} = k_\textrm{there}$. In addition, the vector states contained in the supermultiplets with dimensions (\ref{eq:SpecGRAVG2IIA})--(\ref{eq:SpecVECG2IIA}) reproduce the masses given in table 15 of \cite{Varela:2020wty} up to KK level $n=2$. Like their counterparts (\ref{eq:SpecGRAVG2D11})--(\ref{eq:SpecCHIRALG2D11}) for the $\cN=1$ G$_2$ spectrum in $D=11$, all the dimensions (\ref{eq:SpecGRAVG2IIA})--(\ref{eq:SpecCHIRALG2IIA}) for the type IIA spectrum also conform to the generic expression (\ref{eq:E0generic}) brought to the introduction, now with $d=6$, $\alpha = \tfrac56$, $\beta = \tfrac54$ therein. This is straightforward to see for the GINO, (M)VEC and CHIRAL dimensions (\ref{eq:SpecGINOG2IIA})--(\ref{eq:SpecCHIRALG2IIA}), by using the relevant values of $s_0$ from table \ref{tab:OSp(4|1) supermultiplets}. To see that the (M)GRAV dimension (\ref{eq:SpecGRAVG2IIA}) can be also rewritten as in (\ref{eq:E0generic}), 
\begin{equation}
E_0 = 1 + \sqrt{ \tfrac94 + \tfrac56 n (n+5) - \tfrac54 \, {\cal C}_2 (p,q) } \; , 
\end{equation}
note from tables \ref{tab:multipletsatlevel0G2IIA}--\ref{tab:multipletsatlevel3G2IIA} that, at KK level $n$, there is a unique (M)GRAV supermultiplet that occurs with G$_2$ quantum numbers $p=n$, $q=0$, and that ${\cal C}_2( n,0) = \tfrac13 n (n+5)$ by (\ref{eq:G2CasimirEigenv}).

Finally, like for the $\cN=1$ G$_2$ solution of $D=11$ supergravity, the spectrum of individual KK scalar states for the $\cN=1$ G$_2$ solution of type IIA can be deduced from the complete supersymmetric spectrum that we have presented in this section. Table \ref{tab:N=1G2ISO7Scalars} compiles the result for all scalars in the spectrum with conformal dimension $\Delta \leq 3$, following the same layout as table \ref{tab:N=1G2SO(8)calars}. In this case, all scalars with $\Delta \leq 3$ arise at KK levels $n=0,1,2$. Each of these KK levels also contain scalars with $\Delta >3$. At KK levels $n \geq 3$, all scalars have dimensions  $\Delta >3$.

 \begin{table}%[H]

\centering

\scriptsize{

\begin{tabular}{llccclcl}
\hline
\multicolumn{2}{c}{$\Delta$}  && $n $  &&   $\textrm{G}_2  $  && $\cN=1$ supermultiplet
            \\ \hline
 %
 %
 %
 %
 % 
%
% \hline
 %
 %
$ 1 + \tfrac{\sqrt{6}}{6} $ & $1.41$ & &  0   &&  $[2,0]     $   && CHIRAL $\left( 1 + \tfrac{\sqrt{6}}{6}   \right)  $  \\[5pt] 
 %
%
% \hline
 %
 %
$ 2 $ & $2.00$ & & 1   &&  $[3,0]    $   && CHIRAL $\left( 2  \right)  $  \\[5pt] 
 %
%
% \hline
 %
$ 2 + \tfrac{\sqrt{6}}{6} $ & $2.41$ & &  0   &&  $[2,0]     $   && CHIRAL $\left( 1 + \tfrac{\sqrt{6}}{6}   \right)  $  \\[5pt] 
 %
%
%
 % \hline
 %
 %
$ 1 + \tfrac{2\sqrt{6}}{3} $ & $2.63$ & & 2   &&  $[4,0]  $   && CHIRAL $\left( 1 + \tfrac{2\sqrt{6}}{3} \right) $ \\[5pt] 
 %
%
 % \hline
 %
 %
$ \tfrac32 + \sqrt{\tfrac32} $ & $2.72$ & & 1   &&  $[1,1]  $   && VEC $\left( 1 + \sqrt{\tfrac32} \right) $ \\[5pt] 
 %
%
%
 % \hline
 %
 %
%
%
%
$3$ & $3.00$ & &  1   &&  $[3,0]  $   && CHIRAL $\left( 2 \right)   $   \\[5pt] 
 \hline
\end{tabular}

}\normalsize

\caption{\scriptsize{All KK scalars with dimension $\Delta \leq 3$ around the $\cN=1$ SU(3)-invariant AdS$_4$ solution of type IIA.  }\normalsize}
\label{tab:N=1G2ISO7Scalars}
\end{table}

%%%%%%%%%%%%%%%

\subsection{Spectrum of the $\cN=1$ SU(3)-invariant solution } \label{sec:N=1SU3SpectrumIIA}

%%%%%%%%%%%%%%%

We finally turn to the $\cN=1$ AdS$_4$ solution of type IIA with SU(3) residual symmetry. A critical point of maximal supergravity with dyonic gaugings \cite{Dall'Agata:2012bb,Dall'Agata:2014ita,Inverso:2015viq} with this residual (super)symmetry was first reported in \cite{Borghese:2012zs}. This vacuum was more precisely identified as a critical point of the dyonic ISO(7) gauging in \cite{Guarino:2015qaa}. The resulting ten-dimensional AdS$_4$ solution was constructed in \cite{Varela:2015uca} using the uplifting formulae of \cite{Guarino:2015jca,Guarino:2015vca}. The massive type IIA solution, (4.4), (4.5) of \cite{Varela:2015uca},   features a warped product of AdS$_4$ with a topological $S^6$. The latter is equipped with a cohomogeneity-one metric. This may be seen as a deformation of the usual sine-cone metric over $S^5$, where the U(1) Hopf fibre of the latter is inhomogeneously squashed against the $\mathbb{CP}_2$ base, so that the isometry is $\textrm{SU}(3) \times \textrm{U}(1)$. The $S^5$ is endowed with its canonical Sasaki-Einstein structure, along whose forms take values the type IIA fluxes. The symmetry of the full solution is thus reduced to SU(3). The $\cN=1$ supersymmetry is captured by a type of $G$-structure discussed in \cite{Lust:2009zb}. 

Like in the previous cases, the KK spectrum of this solution is known partially. The bosonic spectrum at KK level $n=0$ was given in \cite{Guarino:2015qaa}. The $n=0$ fermion spectrum follows by supersymmetry from its bosonic counterpart, and has been explicitly given in \cite{Bobev:2020qev}. At higher KK levels, only the spectra of KK gravitons \cite{Pang:2017omp} and KK vectors \cite{Varela:2020wty} are known. In the present paper, we have determined the spectrum of KK gravitini and spin-$1/2$ fermions diagonalising the mass matrices of section \ref{sec:KKFermionMassMat}. These new and previous results allow us to determine the complete supersymmetric KK spectrum above this $\cN=1$ AdS$_4$ solution. 
 
The complete KK spectrum in this case comes in representations of $\textrm{OSp}(4|1) \times \textrm{SU}(3)$. Other than this, the allocation of the spectra into supermultiplets proceeds as in section \ref{sec:N=1G2SpectrumIIA}. The conformal dimensions are again left undertermined by the group theory, but these can be brought in from \cite{Pang:2017omp,Varela:2020wty} and from the explicit calculation of the fermionic sector described above. The result up to KK level $n=3$ is summarised in tables \ref{tab:multipletsatlevel0}--\ref{tab:multipletsatlevel3} below. Again, the dimension $(E_0)$ is shown next to each supermultiplet. An entry of the form $m \times (E_0)$ indicates that there are $m$ such supermultiplets, with the label $m$ omitted when $m=1$. Recall that the representation $[p,q]$ of SU(3) with $p \neq q$ is complex, and its conjugate is $[q,p]$. In order to avoid repetition, the supermultiplets with SU(3) quantum numbers $[p,q]$ with $q>p$ are simply indicated as the complex conjugates of those with quantum numbers $[q,p]$. Supermultiplets in conjugate representations have the same conformal dimension $E_0$.  For example, from table \ref{tab:multipletsatlevel1}, the KK spectrum includes $\overline{\bm{6}}$ CHIRAL multiplets with dimension $E_0 = \frac{10}{3}$ at KK level $n=1$.

\vspace{50pt}

%Multiplets at level 0

\begin{table}[H]
\begin{center}
{\footnotesize
\begin{tabular}{|p{35mm}|p{30mm}|p{30mm}|} 					\hline
$[0,0]$ 				& 	$[0,1]$ 				& 	$[0,2]$ 			\\[1pt]
MGRAV $\left( \tfrac52 \right)$  			& conj.~to $[1,0]$ 	& conj.~to $[2,0]$ 	\\
GINO  $\left( 3 \right)$ 		& & 				\\
CHIRAL  $ 2 \times \left( 1+ \sqrt{6} \right) $  		& & 					\\[5pt] \hline
$[1,0]$ 				& 	$[1,1]$ 			\\[1pt]
GINO $\left( \tfrac{7}{3} \right) $ & MVEC $\left( \tfrac32 \right)$ 			\\[5pt] 
VEC $\left( 1+ \tfrac{\sqrt{109}}{6} \right) $ & CHIRAL $\left( 2 \right)$ 			\\[5pt] \cline{1-2}
$[2,0]$ 			\\[1pt]
CHIRAL $\left( \tfrac{5}{3} \right)$ 	\\[5pt] \cline{1-1}
\end{tabular}
}
\caption{\footnotesize{Supermultiplets at KK level $n=0$ for the $\cN=1$ SU(3)-invariant AdS$_4$ solution of type IIA.}\normalsize}
\label{tab:multipletsatlevel0}
\end{center}
\end{table}

\vspace{50pt}

%Multiplets at level 1

\begin{table}[H]
\begin{center}
{\footnotesize
\begin{tabular}{|p{35mm}|p{35mm}|p{35mm}|p{35mm}|} \hline
$[0,0]$ 							& 	$[0,1]$ 				& 	$[0,2]$ 			& 	$[0,3]$ \\[1pt]
GRAV $\left( 1+\tfrac{\sqrt{29}}{2} \right)$			& conj.~to $[1,0]$ 	& conj.~to $[2,0]$ 	& conj.~to $[3,0]$ \\
GINO $2 \times \left( 4 \right)$	&        	&      	& \\
VEC $2 \times \left( 1+\tfrac{\sqrt{41}}{2} \right)$	&        	&      	& \\
CHIRAL $2 \times \left( 1+ \sqrt{11} \right)$	&        	&      	&  \\[5pt] \hline
$[1,0]$ 					& 	$[1,1]$ 					& 	$[1,2]$ 			\\[1pt]
GRAV $\left( 1+\tfrac{\sqrt{181}}{6} \right)$	& GINO $2 \times \left( 3 \right)$	& conj.~to $[2,1]$ 		\\
GINO  $2 \times \left( 1+\tfrac{\sqrt{61}}{3} \right)$		& VEC $3\times \left( 1+\tfrac{\sqrt{21}}{2} \right)$ 		&     	\\
VEC $ 3\times  \left(  \tfrac{23}{6} \right) $			&         CHIRAL $2 \times \left( 1+ \sqrt{6} \right)$ 						&                    	\\
CHIRAL $ 3 \times \left( 1+\tfrac{\sqrt{79}}{3} \right)$			&           						&                    			\\[5pt]  \cline{1-3}
$[2,0]$ 			& 	$[2,1]$ 				\\[1pt]
GINO $\left( 1+\tfrac{\sqrt{31}}{3} \right) $ 	& VEC  $\left( \tfrac{13}{6} \right)$	\\
VEC $2 \times \left( \tfrac{19}{6} \right) $ 	& CHIRAL $ \left( 1+\tfrac{\sqrt{19}}{3} \right)$		\\
CHIRAL $ \left( \tfrac{10}{3} \right) $  	&  	\\[5pt] \cline{1-2}
$[3,0]$ \\[1pt]
CHIRAL $ \left( 2 \right) $  \\[5pt] \cline{1-1}
\end{tabular}
}
\caption{\footnotesize{Supermultiplets at KK level $n=1$ for the $\cN=1$ SU(3)-invariant AdS$_4$ solution of type IIA.}}
\label{tab:multipletsatlevel1}
\end{center}
\end{table}

\vspace{40pt}

%Multiplets at level 2

\begin{table}[H]
\begin{center}
{\footnotesize
\begin{tabular}{|p{35mm}|p{35mm}|p{35mm}|p{20mm}|p{20mm}|} \hline
$[0,0]$ 								& 	$[0,1]$ 			& 	$[0,2]$ 			& 	$[0,3]$			& 	$[0,4]$ 		\\[1pt]
GRAV $\left( 1+ \tfrac{\sqrt{501}}{6} \right)$								& 	conj. to [1,0] 			 	& 	conj. to [2,0]				&  conj. to [3,0]					&		conj. to [4,0]		\\[5pt]
GINO $2 \times \left( 1+ \sqrt{\tfrac{47}{3}} \right)$				& 		 	& 			& 			&	\\[5pt]
VEC $3 \times \left( 1+ \tfrac{\sqrt{609}}{6} \right)$				& 		 	& 			& 			&	\\[5pt]
CHIRAL $4 \times \left( 1+ \sqrt{\tfrac{53}{3}} \right)$				& 		 	& 			& 			&		\\[5pt] \hline
$[1,0]$ 										& 	$[1,1]$ 						& 	$[1,2]$ 			&	$[1,3]$			\\[1pt]
GRAV $\left( 1+ \tfrac{\sqrt{421}}{6} \right)$				& 	GRAV $\left( 1 + \tfrac{\sqrt{321}}{6} \right)$					& 					&					\\[4pt]
GINO $4 \times \left( \tfrac{14}{3} \right) $		& 				GINO $4 \times \left( 1 + \tfrac{4\sqrt{6}}{3} \right)$	&      conj. to [2,1]		&	conj. to [3,1]		\\[5pt]
VEC $6 \times \left(  \tfrac{29}{6} \right)$ 					& 	VEC $ 7 \times \left( 1+ \tfrac{\sqrt{429}}{6} \right)$		&      					&			\\[5pt]
CHIRAL $4 \times \left( 1+ \tfrac{\sqrt{139}}{3} \right)$ 		& 	CHIRAL $ 6 \times \left( 1+ \sqrt{\tfrac{38}{3}} \right)$		&      					&					\\[8pt] \cline{1-4}
$[2,0]$ 										&	$[2,1]$ 										&	$[2,2]$		\\[1pt]
GRAV $\left( 1+ \tfrac{\sqrt{301}}{6}  \right)$			&	GINO  $2 \times \left( 1+ \tfrac{\sqrt{61}}{3} \right) $ 		&	VEC $\left( 1 +\tfrac{ \sqrt{129}}{6} \right)$  \\
GINO $2\times \left( 1 + \tfrac{\sqrt{91}}{3} \right)$	 	&	VEC $4 \times \left( \tfrac{23}{6}  \right)$				&  	CHIRAL $2 \times \left( 1 + \sqrt{\tfrac{13}{3}}  \right)$								 \\
VEC $4 \times \left(  1 + \tfrac{\sqrt{409}}{6}  \right)$		 &	CHIRAL $3 \times \left( 1 + \tfrac{\sqrt{79}}{3}  \right)$	&   \\[4pt]
CHIRAL $5 \times \left( 1+ \tfrac{\sqrt{109}}{3} \right)$	&						&   \\[6pt] \cline{1-3}
$[3,0]$ 										&	$[3,1]$			\\[1pt]
GINO $\left( 1 + \sqrt{\tfrac{17}{3}} \right)$	 			&	VEC $\left( 1 + \tfrac{\sqrt{109}}{6} \right) $\\[1pt]
VEC $2 \times \left( 1 + \tfrac{\sqrt{249}}{6}  \right)$	 	&	CHIRAL $\left( 1 + \tfrac{\sqrt{34}}{3} \right)$\\[1pt]
CHIRAL $ \left(1 + \sqrt{\tfrac{23}{3}}  \right)$	 		&		\\[8pt]  \cline{1-2}
$[4,0]$ 			\\[1pt]
CHIRAL $\left( 1 + \tfrac{\sqrt{19}}{3}  \right)$	 	\\[5pt] \cline{1-1}
\end{tabular}
}
\caption{\footnotesize{Supermultiplets at KK level $n=2$ for the $\cN=1$ SU(3)-invariant AdS$_4$ solution of type IIA.}\normalsize}
\label{tab:multipletsatlevel2}
\end{center}
\end{table}

%Multiplets at level 3

\begin{table}[H]
\begin{center}
\begin{sideways}
{\scriptsize
\begin{tabular}{|l|l|l|l|l|l|} \hline
$[0,0]$ & $[0,1]$ & $[0,2]$ & $[0,3]$ & $[0,4]$ & $[0,5]$ \\
GRAV $\left( 1 +\tfrac{\sqrt{89}}{2} \right)$ & conj. to $[1,0]$ & conj. to $[2,0]$ & conj. to $[3,0]$ & conj. to $[4,0]$ & conj. to $[5,0]$ \\[1pt]
GINO $2 \times \left( 1+2\sqrt{6} \right)$ & & & & &  \\[1pt]
VEC $3 \times \left( 1 +\tfrac{\sqrt{101}}{2} \right)$ & & & & &  \\[1pt]
CHIRAL $4 \times \left( 1+\sqrt{26} \right)$ & & & & &     \\[5pt] \hline
$[1,0]$ & $[1,1]$ & $[1,2]$ & $[1,3]$ & $[1,4]$ \\[1pt]
GRAV $\left( 1 +\tfrac{\sqrt{721}}{6} \right)$   				& GRAV $\left( 1 +\tfrac{\sqrt{69}}{2} \right)$ & conj. to $[2,1]$ & conj. to $[3,1]$ & conj. to $[4,1]$ \\[7pt]
GINO $4 \times \left( \tfrac{17}{3} \right)$  					& GINO $6 \times \left( 1+ \sqrt{19} \right)$  	 & & & \\
VEC $7 \times \left( 1 + \tfrac{\sqrt{829}}{6} \right)$			 & VEC $11 \times \left( \tfrac{11}{2} \right)$  & & &\\
CHIRAL $6 \times \left( 1 + \tfrac{\sqrt{214}}{3} \right)$ 			& CHIRAL $8 \times \left( 1+ \sqrt{21} \right)$  & & & \\[8pt] \cline{1-5}
$[2,0]$ & $[2,1]$ & $[2,2]$ & $[2,3]$ \\[1pt]
GRAV $\left( 1 +\tfrac{\sqrt{601}}{6} \right)$  				& GRAV $\left( 1 +\tfrac{\sqrt{481}}{6} \right)$    			& GINO $2 \times \left( 1 +\tfrac{4\sqrt{6}}{3} \right)$ & conj. to $[3,2]$ \\[8pt]
GINO $4 \times \left( 1 +\tfrac{\sqrt{166}}{3} \right)$  			& GINO $4 \times \left( 1 +\tfrac{2\sqrt{34}}{3} \right)$    		 & VEC $5  \times \left( 1 + \tfrac{\sqrt{429}}{6} \right)$   & \\[10pt]
VEC $7 \times \left( 1 + \tfrac{\sqrt{709}}{6} \right)$     		& VEC $8  \times \left( 1 + \tfrac{\sqrt{589}}{6} \right)$      		& CHIRAL $4 \times \left( 1 + \tfrac{\sqrt{114}}{3} \right)$ & \\[10pt]
CHIRAL $6 \times \left( 1 + \tfrac{2\sqrt{46}}{3} \right)$     		& CHIRAL $8 \times \left( 1 + \tfrac{\sqrt{154}}{3} \right)$    	& & \\[8pt]\cline{1-4}
$[3,0]$ & $[3,1]$ & $[3,2]$ \\[1pt]
GRAV $\left( \tfrac92 \right)$ 							& GINO $2 \times \left( 1 +\tfrac{\sqrt{91}}{3} \right)$ 				&  VEC $ \left( 1 + \tfrac{\sqrt{229}}{6} \right)$ \\[4pt]
GINO $2\times \left( 1 +\sqrt{14} \right)$ 					& VEC $4 \times \left( 1 +\tfrac{\sqrt{409}}{6} \right)$ 				& CHIRAL $2 \times \left( \tfrac{11}{3} \right)$  	  \\[4pt]
VEC $4\times \left( 1 + \tfrac{\sqrt{61}}{2} \right)$  			  &  CHIRAL $3 \times \left( 1 +\tfrac{\sqrt{109}}{3} \right)$ 		&\\[4pt]
CHIRAL $5\times \left( 5 \right)$    						&     & \\[8pt] \cline{1-3}
$[4,0]$ & $[4,1]$ \\[1pt]
GINO $\left( 1 + \tfrac{2\sqrt{19}}{3} \right)$  				&  VEC $\left( 1 +\tfrac{\sqrt{21}}{2} \right)$  \\
VEC $2\times \left( 1 + \tfrac{\sqrt{349}}{6} \right)$  			&  CHIRAL $ \left( 1 +\sqrt{6} \right)$  \\
CHIRAL $\left( 1 + \tfrac{\sqrt{94}}{3} \right)$  				& 							  \\[5pt] \cline{1-2}
$[5,0]$ \\[1pt]
CHIRAL $\left( 1 + \tfrac{\sqrt{34}}{3} \right)$   \\[8pt] \cline{1-1}
\end{tabular}
}
\end{sideways}
\caption{\footnotesize{Supermultiplets at KK level $n=3$ for the $\cN=1$ SU(3)-invariant AdS$_4$ solution of type IIA.}\normalsize}
\label{tab:multipletsatlevel3}
\end{center}

\end{table}

\newpage

The $\textrm{OSp}(4|1)$ representations at level $n=0$ have recently appeared in \cite{Bobev:2020qev}, and table \ref{tab:multipletsatlevel0} matches their results. As expected, the $n=0$ spectrum contains a singlet MGRAV and $\bm{8}$ MVECs: the former is the $\cN=1$ supergravity multiplet and the latter contains the vectors that gauge the residual SU(3) symmetry. More generally, like in the previous $\cN=1$ cases discussed in sections \ref{sec:N=1G2SpectrumD=11} and 
\ref{sec:N=1G2SpectrumIIA}, closed form expressions may be given for the conformal dimensions at all levels $n \geq 0$. From tables \ref{tab:multipletsatlevel0}--\ref{tab:multipletsatlevel3}, the conformal dimension $E_0$ of each type of OSp$(4|1)$ supermultiplet at KK level $n$ with SU(3) Dynkin labels $[p,q]$ turns out to be given by 
{\setlength\arraycolsep{2pt}
\begin{eqnarray} \label{eq:SpecGRAVSU3IIA}
 \textrm{(M)GRAV } & : & E_0 = 1 + \sqrt{ \tfrac94 + \tfrac56 n (n+5) - \tfrac53 \, {\cal C}_2 (p,q) } \; ,   \hspace{10pt} \\[5pt]
\label{eq:SpecGINOSU3IIA}
 \textrm{GINO}  & : & E_0 = 1 + \sqrt{ 4 + \tfrac56 n (n+5) - \tfrac53 \, {\cal C}_2 (p,q) } \; ,   \hspace{10pt} \\[5pt]
\label{eq:SpecVECSU3IIA}
 \textrm{(M)VEC} & : & E_0 = 1 + \sqrt{ \tfrac{21}{4} + \tfrac56 n (n+5) - \tfrac53 \, {\cal C}_2 (p,q) } \; ,   \hspace{10pt} \\[5pt]
\label{eq:SpecCHIRALSU3IIA}
 \textrm{CHIRAL} & : & E_0 = 1 + \sqrt{ 6 + \tfrac56 n (n+5) - \tfrac53 \, {\cal C}_2 (p,q) } \; .
\end{eqnarray}
}Here, like in (\ref{eq:SpecGRAVG2IIA})--(\ref{eq:SpecCHIRALG2IIA}) for the type IIA $\cN=1$ G$_2$ solution, $n(n+5)$ are the eigenvalues of the scalar Laplacian on $S^6$, but now
\begin{equation} \label{eq:SU3CasimirEigenv}
{\cal C}_2( p,q) \equiv  \tfrac13 \left[  p (p+3) + q (q+3)  + pq  \right]  
\end{equation}
are the eigenvalues of the quadratic Casimir operator of SU(3) in the $[p,q]$ representation, normalised as indicated in footnote \ref{fn:NormCas}. The (M)GRAV dimensions (\ref{eq:SpecGRAVSU3IIA}) match the masses for the  graviton states, given in (3.1) of \cite{Pang:2017omp}, with $k_\textrm{there} = n_\textrm{here}$, $\ell_\textrm{there} = p_\textrm{here} + q_\textrm{here}$, $p_\textrm{there} = p_\textrm{here}$. The individual vector masses that follow from (\ref{eq:SpecGRAVSU3IIA})--(\ref{eq:SpecVECSU3IIA}) match table 15 of \cite{Varela:2020wty} up to KK level $2$. As for the previous cases, the dimensions (\ref{eq:SpecGRAVSU3IIA})--(\ref{eq:SpecCHIRALSU3IIA}) can also be written compactly as (\ref{eq:E0generic}) of the introduction with $d=6$, $\alpha = \tfrac56$ and $\beta = \tfrac53$.

 \begin{table}%[H]

\centering

\scriptsize{

\begin{tabular}{llccclcl}
\hline
\multicolumn{2}{c}{$\Delta$}  && $n $  &&   $\textrm{SU}(3)  $  && $\cN=1$ supermultiplet
            \\ \hline
 $\tfrac53$ & $1.67$  &&  0   &&  $[2,0] + \textrm{c.c.} $   && CHIRAL  $\left( \tfrac{5}{3} \right) + \textrm{c.c.}  $ \\[5pt] 
%
% \hline
 %
 %
$2$ & $2.00$ & &  0   &&  $[1,1]  $   && CHIRAL $\left( 2 \right)   $   \\[5pt] 
 %
%
% \hline
 %
 %
 $2$ & $2.00$  &&  1   &&  $[3,0]  + \textrm{c.c.} $   && CHIRAL $\left( 2 \right) $  \\[5pt] 
%
% \hline
 %
 %
$ 1 + \tfrac{\sqrt{19}}{3} $ & $2.45$ & &  1   &&  $[2,1]  + \textrm{c.c.}     $   && CHIRAL $\left( 1 + \tfrac{\sqrt{19}}{3}  \right) + \textrm{c.c.}  $  \\[5pt] 
 %
%
% \hline
 %
 %
$ 1 + \tfrac{\sqrt{19}}{3} $ & $2.45$ & &  2   &&  $[4,0]  + \textrm{c.c.}     $   && CHIRAL $\left( 1 + \tfrac{\sqrt{19}}{3}  \right) + \textrm{c.c.}  $  \\[5pt] 
 %
%
% \hline
 %
 %
 $\tfrac83$ & $2.67$  &&  0   &&  $[2,0]  + \textrm{c.c.} $   &&  CHIRAL $\left( \tfrac{5}{3} \right) + \textrm{c.c.}  $   \\[5pt] 
%
% \hline
 %
 %
 $\tfrac83$ & $2.67$  &&  1   &&  $[2,1]  + \textrm{c.c.} $   && VEC $\left( \tfrac{13}{6} \right) + \textrm{c.c.}  $ \\[5pt] 
%
%
 % \hline
 %
 %
$ 1 + \tfrac{\sqrt{34}}{3} $ & $2.94$ & & 2   &&  $[3,1]+ \textrm{c.c.}  $   && CHIRAL $\left( 1 + \tfrac{\sqrt{34}}{3} \right) $ \\[5pt] 
 %
%
 % \hline
 %
 %
$ 1 + \tfrac{\sqrt{34}}{3} $ & $2.94$ & & 3   &&  $[5,0]+ \textrm{c.c.}  $   && CHIRAL $\left( 1 + \tfrac{\sqrt{34}}{3} \right) $ \\[5pt] 
$3$ & $3.00$ & &  0   &&  $[1,1]  $   && CHIRAL $\left( 2 \right)   $   \\[5pt] 
 %
%
% \hline
 %
 %
 $3$ & $3.00$  &&  1   &&  $[3,0]  + \textrm{c.c.} $   && CHIRAL $\left( 2 \right) $  \\[5pt] 
 \hline
\end{tabular}

}\normalsize

\caption{\scriptsize{All KK scalars with dimension $\Delta \leq 3$ around the $\cN=1$ SU(3)-invariant AdS$_4$ solution of type IIA.  }\normalsize}
\label{tab:N=1SU3ISO7Scalars}
\end{table}

Again like in the previous cases, the KK scalar spectrum for the $\cN=1$ SU(3) type IIA solution follows from the complete supersymmetric spectrum. All KK scalars with dimensions $\Delta \leq 3$ are summarised in table \ref{tab:N=1SU3ISO7Scalars}, following the same notation and conventions as tables \ref{tab:N=1G2SO(8)calars} and \ref{tab:N=1G2ISO7Scalars}. As already noted in \cite{Guarino:2015qaa}, massless ($\Delta = 3$) scalars, in the adjoint of SU(3), already appear at KK level $n=0$. This is an unusual feature for this type of AdS$_4$ solutions, for which massless scalars tend to appear at higher KK levels. The massless scalar spectrum is completed at level $n=1$ with $\bm{10} + \overline{\bm{10}}$ more scalars. All scalars with dimension $\Delta <3$ appear at KK levels $n = 0, 1, 2 , 3$. These levels also contain scalars with $\Delta >3$. For KK levels $n \geq 4$, all scalars have $\Delta >3$.

%%%%%%%%%%%%%%%

%\section{Spectrum of the $\cN=0$ G$_2$-invariant solution of type IIA} \label{sec:N=0G2SpectrumIIA}

\section{Complete non-supersymmetric KK spectra} \label{sec:N=0G2SpectrumIIA}

%%%%%%%%%%%%%%%

Within the class of $D=11$ and type IIA AdS$_4$ solutions with at least SU(3) symmetry that we are considering, there are non-supersymmetric solutions besides the $\cN=1$ cases discussed in sections \ref{sec:N=1Spectrum11D} and \ref{sec:N=1Spectrum10D}. These include solutions with symmetry SO(7)$_v$ \cite{deWit:1984va}, SO(7)$_c$ \cite{Englert:1982vs} and SU$(4)_c$ \cite{Pope:1984bd} in $D=11$, and SO(7) \cite{DallAgata:2011aa,Varela:2015uca,Romans:1985tz}, SO(6) \cite{DallAgata:2011aa,Varela:2015uca} and G$_2$ \cite{Borghese:2012qm,Varela:2015uca,Lust:2008zd} in type IIA. In addition, there are two type IIA  solutions with SU(3) symmetry which are only known numerically \cite{Guarino:2015qaa,Varela:2015uca} and will be excluded from our discussion. See respectively \cite{Larios:2019kbw} and \cite{Varela:2015uca} for these eleven- and ten-dimensional AdS$_4$ solutions in our conventions. In this section we turn to address the KK spectrum for these solutions. 

The state-of-the-art for their KK spectra is the following. For all of these, the bosonic \cite{Bobev:2010ib,DallAgata:2011aa,Borghese:2012qm} and fermionic \cite{Comsa:2019rcz,Bobev:2020qev} spectra at KK level $n=0$ are known. At higher KK levels only the bosonic spectra are known, either partially or completely. The spectra of KK gravitons \cite{Dimmitt:2019qla,Pang:2017omp} and vectors \cite{Varela:2020wty} are known and, for the G$_2$ solution of type IIA, also the KK scalar spectrum is known \cite{Guarino:2020flh}. Thus, the bosonic KK spectrum of the G$_2$ solution is completely known. For the other solutions, the bosonic spectrum is known short of the KK scalars. In this section, we will give the fermionic spectra for all these solutions, thereby completing the KK spectrum for the type IIA G$_2$ solution. For all other solutions, we will conjecture a formula for the KK scalar masses based on strong plausibility arguments. This will effectively complete their KK spectra as well.

Using our fermionic KK mass matrices, we have computed the KK gravitino and spin-$1/2$ fermion spectra for all these solutions. We have recorded the gravitino mass eigenvalues up to KK level $n=2$ in tables \ref{tab:KKGravitiniSO8} and \ref{tab:KKGravitiniISO7} of appendix \ref{sec:SpecificSpectra}. For the G$_2$ solution, also the spin-$1/2$ fermion spectrum is tabulated in table  \ref{tab:KKSpin1/2G2N=0} of the appendix, again up to second KK level. The fields of different spin present in these spectra are  organised KK level by KK level in representations of the residual symmetry group $G$, with $G= \textrm{SO}(7)$ or $G= \textrm{SO}(6) \sim \textrm{SU} (4) $, with Dynkin labels $[p,q,r]$, or $G= \textrm{G}_2$, with Dynkin labels $[p,q]$. For the $D=11$ solutions, these representations branch from the SO(8) representations of the spectrum at the $\cN=8$ SO(8) point \cite{Englert:1983rn} (see also {\it e.g.}~table 2 of \cite{Klebanov:2008vq} for a summary). For the type IIA solutions, the representations split instead from the SO(7) representations given in table 1 of \cite{Varela:2020wty}. For the G$_2$ solution, we have recorded the G$_2$ representation content of the spectrum up to KK level $n=3$ in tables \ref{tab:statesatlevel0G2IIA}--\ref{tab:statesatlevel3G2IIA} below. Together with this group theory analysis and the previously known results for the bosonic sector \cite{Borghese:2012qm,Pang:2017omp,Varela:2020wty,Guarino:2020flh} of the KK spectrum, our new fermionic results finally complete the spectrum of the non-supersymmetric G$_2$-invariant solution of type IIA. Further, closed-form formulae can be given for the masses at all KK levels, for this and the other solutions, as we will see momentarily.

 \begin{table}[]

%\centering

\resizebox{\textwidth}{!}{

\begin{tabular}{|l|l|l l l l l |}
\hline
$D=11$ sol.               & $s$                  &               &&    $ L^2M^2$  &   &           \\ \hline \hline
\multirow{6.2}{*}{$\cN=0$, SO$(7)_v$} & $2$   & &  &$\frac{3}{4}n(n+6)$& $-$ &$\frac{6}{5} \, \mathcal{C}_2 (p,q,r)$            \\[4pt]  
\cline{2-7} %\cline{7} 
& $ \frac{3}{2}$ &   $\frac{9}{2}$&+ & $\frac{3}{4} \, n(n+6)$ & $-$ &$\frac{6}{5} \, \mathcal{C}_2 (p,q,r)$
 \\[4pt]  
\cline{2-7} %\cline{7} 
& $ 1$ &    $6$&+ & $\frac{3}{4} \, n(n+6)$ &  $-$ &$\frac{6}{5} \, \mathcal{C}_2 (p,q,r)$ \\[4pt]  
\cline{2-7} %\cline{7} 
& $ \frac{1}{2}$ &    $\frac{15}{2}$&+& $\frac{3}{4} \, n(n+6)$ &  $-$ &$\frac{6}{5} \, \mathcal{C}_2 (p,q,r)$ \\[4pt]
\cline{2-7}
& $0^{(*)}$ &    6&+ & $\frac{3}{4} \, n(n+6)$ & $-$ &$\frac{6}{5} \, \mathcal{C}_2 (p,q,r)$ 
\\[4pt]  \hline \hline
\multirow{6.2}{*}{$\cN=0$, SO$(7)_c$} & $2$    &  && $ \frac{3}{4} \, n(n+6)$ & $-$ &$\frac{6}{5} \, \mathcal{C}_2 (p,q,r)$           \\[4pt]  
\cline{2-7} %\cline{7} 
& $ \frac{3}{2}$ & $\frac{9}{2}$&+&$\frac{3}{4} \, n(n+6)$& $-$ &$\frac{6}{5} \, \mathcal{C}_2 (p,q,r)$ 
 \\[4pt]  
\cline{2-7} %\cline{7} 
& $ 1$ &    6&+&$\frac{3}{4} \, n(n+6)$& $-$ &$\frac{6}{5} \, \mathcal{C}_2 (p,q,r)$ 
\\[4pt] \cline{2-7} %\cline{7} 
& $ \frac{1}{2}$ &   $\frac{15}{2}$&+&$\frac{3}{4} \, n(n+6)$& $-$ &$\frac{6}{5} \, \mathcal{C}_2 (p,q,r)$ 
\\[4pt] 
\cline{2-7} %\cline{7} 
& $0^{(*)}$ &   6&+&$\frac{3}{4} \, n(n+6)$& $-$ &$\frac{6}{5} \, \mathcal{C}_2 (p,q,r)$ 
\\[4pt] 
 \hline \hline
\multirow{6.2}{*}{$\cN=0$, SU$(4)_c$} & $2$   && & $\frac{3}{4} \, n(n+6)$& $-$ &$\frac{3}{2} \, \mathcal{C}_2 (p,q,r)$          \\[4pt]  
\cline{2-7} %\cline{7} 
& $ \frac{3}{2}$ & $\frac{9}{2}$&+& $\frac{3}{4} \, n(n+6)$& $-$ &$\frac{3}{2} \, \mathcal{C}_2 (p,q,r)$
 \\[4pt]  
\cline{2-7} %\cline{7} 
& $ 1$ & 6&+&$\frac{3}{4} \, n(n+6)$& $-$ &$\frac{3}{2} \, \mathcal{C}_2 (p,q,r)$

\\[4pt] 
\cline{2-7} %\cline{7} 
& $ \frac{1}{2}$ & $\frac{15}{2}$ &+&$\frac{3}{4} \, n(n+6)$& $-$ &$\frac{3}{2} \, \mathcal{C}_2 (p,q,r)$

\\[4pt]
\cline{2-7} %\cline{7} 
& $0^{(*)}$ & 6&+&$\frac{3}{4} \, n(n+6)$& $-$ &$\frac{3}{2} \, \mathcal{C}_2 (p,q,r)$

\\[4pt] \hline 
\end{tabular}

\quad

\begin{tabular}{|l|l|l l l l l |}
\hline
IIA sol.                & $s$                  &  && $L^2M^2$  &&                   \\ \hline \hline
\multirow{6.2}{*}{$\cN=0$, SO$(7)$} & $2$  &  & &$ n(n+5)$& $-$ &$\frac{6}{5} \, \mathcal{C}_2 (p,q,r)$            \\[4pt]  
\cline{2-7} %\cline{7} 
& $ \frac{3}{2}$ &   $\frac{9}{2}$&+&$n(n+5)$& $-$ &$\frac{6}{5} \, \mathcal{C}_2 (p,q,r)$
 \\[4pt]  
\cline{2-7} %\cline{7} 
& $ 1$ &    6&+&$n(n+5)$& $-$ &$\frac{6}{5} \, \mathcal{C}_2 (p,q,r)$ \\[4pt]  
\cline{2-7} %\cline{7} 
& $ \frac{1}{2}$ &    $\frac{15}{2}$&+&$n(n+5)$& $-$ &$\frac{6}{5} \, \mathcal{C}_2 (p,q,r)$ \\[4pt]
\cline{2-7}
& $0^{(*)}$ &    6&+&$n(n+5)$& $-$ &$\frac{6}{5} \, \mathcal{C}_2 (p,q,r)$ 
\\[4pt]  \hline \hline
\multirow{6.2}{*}{$\cN=0$, SO$(6)$} & $2$    & &&  $ n(n+5)$& $-$ &$\frac{3}{2} \, \mathcal{C}_2 (p,q,r)$           \\[4pt]  
\cline{2-7} %\cline{7} 
& $ \frac{3}{2}$ & $\frac{9}{2}$&+&$n(n+5)$& $-$ &$\frac{3}{2} \, \mathcal{C}_2 (p,q,r)$ 
 \\[4pt]  
\cline{2-7} %\cline{7} 
& $ 1$ &   6&+&$n(n+5)$& $-$ &$\frac{3}{2} \, \mathcal{C}_2 (p,q,r)$ 
\\[4pt] \cline{2-7} %\cline{7} 
& $ \frac{1}{2}$ &   $ \frac{15}{2}$&+&$n(n+5)$& $-$ &$\frac{3}{2} \, \mathcal{C}_2 (p,q,r)$ 
\\[4pt] 
\cline{2-7} %\cline{7} 
& $0^{(*)}$ &    6&+&$n(n+5)$& $-$ &$\frac{3}{2} \, \mathcal{C}_2 (p,q,r)$ 
\\[4pt] 
 \hline \hline
\multirow{6.2}{*}{$\cN=0$, G$_2$} & $2$    &&& $n(n+5)$& $-$ &$\frac{3}{2} \, \mathcal{C}_2 (p,q)$          \\[4pt]  
\cline{2-7} %\cline{7} 
& $ \frac{3}{2}$ & $\frac{9}{2}$&+&$n(n+5)$& $-$ &$\frac{3}{2} \, \mathcal{C}_2 (p,q)$
 \\[4pt]  
\cline{2-7} %\cline{7} 
& $ 1$ & 6&+&$n(n+5)$& $-$ &$\frac{3}{2} \, \mathcal{C}_2 (p,q)$

\\[4pt] 
\cline{2-7} %\cline{7} 
& $ \frac{1}{2}$ & $\frac{15}{2}$&+&$n(n+5)$& $-$ &$\frac{3}{2} \, \mathcal{C}_2 (p,q)$

\\[4pt]
\cline{2-7} %\cline{7} 
& $ 0$ & 6&+&$n(n+5)$& $-$ &$\frac{3}{2} \, \mathcal{C}_2 (p,q)$

\\[4pt] \hline 
\end{tabular}

}

\caption{\footnotesize{The complete KK spectra for the analytic non-supersymmetric AdS$_4$ solutions of $D=11$ supergravity (left) and massive type IIA supergravity (right) that respectively uplift from critical points of $D=4$ $\cN=8$ SO(8) and ISO(7) supergravities, with residual symmetry groups larger than SU(3). For each KK field of spin $s$, its squared mass $M^2L^2$ is given at all KK level $n$. The spectra also depend on the quadratic Casimir operators ${\cal C}_2$ specified in the text. The scalar spectra marked with $^{(*)}$ are conjectured.
}\normalsize}
\label{tab:CompleteN=0Spectra}
\end{table}

The spectra of the $\cN=1$ solutions reported in sections \ref{sec:N=1Spectrum11D} and \ref{sec:N=1Spectrum10D} exhibit significant degeneracy, in the sense that all OSp$(4|1)$ supermultiplets of the same type, at the same KK level $n$, and with same SU(3) or G$_2$ quantum numbers $[p,q]$, all have the same conformal dimension $E_0$. However, individual states with the same spin within the same OSp$(4|1)$ supermultiplet necessarily have different masses, as their conformal dimensions must differ by one (see table \ref{tab:OSp(4|1) supermultiplets}). Obviously, this restriction does not affect the non-supersymmetric solutions, as the states do not fill out OSp$(4|1)$ supermultiplets in the first place. In fact, for these $\cN=0$ cases the spectra show an even larger degeneracy: states at the same KK level, with the same spin and the same $G$ quantum numbers, all have the same mass. 

This high degeneracy leads to the existence of closed-form formulae for the mass spectra of these solutions, as already announced above. Closed formulae were given in \cite{Dimmitt:2019qla,Pang:2017omp} for the KK graviton spectra of all these solutions. A mass formula has been similarly given for the KK scalar spectrum of the G$_2$ solution \cite{Guarino:2020flh}. For this solution, the same scalar mass formula has been shown \cite{Guarino:2020flh} to also fit the KK vector spectrum given in \cite{Varela:2020wty} up to KK level $n=2$. Now, we have derived mass formulae for the vector spectra of all other solutions using the data in tables 14 and 15 of \cite{Varela:2020wty}. For our new fermionic results, we have also been able to deduce closed-form mass formulae. Table \ref{tab:CompleteN=0Spectra} summarises all of these. The $D=11$ and IIA graviton spectra listed on the table reproduce the corresponding entries in table 2 of \cite{Dimmitt:2019qla} and equation (3.1) of \cite{Pang:2017omp}, respectively, with the following dictionary of quantum numbers. For the $D=11$ SO$(7)_v$ solution, $n_\textrm{here} = n_\textrm{there}$, $ p_\textrm{here} = k_\textrm{there}$, $q_\textrm{here} =r_\textrm{here} = 0$; for the SO$(7)_c$ solution, $n_\textrm{here} = r_\textrm{here} =  n_\textrm{there}$, $ p_\textrm{here} = q_\textrm{here}=0$; and for the SU$(4)_c$ solution, $n_\textrm{here} = n_\textrm{there} $, $p_\textrm{here} = r_\textrm{there} $,  $q_\textrm{here}=0$, $r_\textrm{here} = n_\textrm{there} - r_\textrm{there} $. For the type IIA gravitons we have, for the SO(7) solution, $n_\textrm{here} =n_\textrm{there}$, $q_\textrm{here} = \ell_\textrm{there} $, $p_\textrm{here} =r_\textrm{here} = 0$; for the SO(6) solution, $n_\textrm{here} =n_\textrm{there}$, $q_\textrm{here} = \ell_\textrm{there} $, $p_\textrm{here} =r_\textrm{here} = 0$; and for the G$_2$ solution, $n_\textrm{here} = p_\textrm{here} = n_\textrm{there}$, $q_\textrm{here} = 0$. The KK scalar and vector formulae for the IIA G$_2$ solution have been imported from (18) of \cite{Guarino:2020flh} with $n_\textrm{here} =\ell_\textrm{there}$, $p_\textrm{here} =n_{1 \, \textrm{there}}$, $q_\textrm{here} =n_{2 \, \textrm{there}}$. All other mass formulae in table \ref{tab:CompleteN=0Spectra} are new. 

As is apparent from the table, a pattern emerges. The squared mass of a state of spin $s$ at KK level $n$, with Dynkin labels $[p,q,r]$ under SO(7) or $\textrm{SU}(4)$ ($[q,p,r]$ for $\textrm{SO} (6) $), or $[p,q]$ under G$_2$, is given by 
\begin{equation} \label{eq:M2L2non-susy}
M^2 L^2 = \gamma_s + \alpha \, n (n+ d-1) - \beta \,  {\cal C}_2 \; .
\end{equation}
Here, $d=7$ in M-theory and $d=6$ in type IIA, as usual. Also, $\gamma_s$ is a constant, the same for all solutions, that only depends on the spin $s$ of the field in question: $\gamma_2 = 0$, $\gamma_{3/2} = \tfrac92$, $\gamma_1 = 6$, $\gamma_{1/2} = \tfrac{15}{2}$ and, for the G$_2$ solution, $\gamma_0 = 6$ as follows from \cite{Guarino:2020flh}. The positive constant $\alpha$ only depends on whether the solution is a solution in M-theory ($\alpha = \tfrac34$) or type IIA ($\alpha = 1$). The positive constant $\beta$ is the same for all solutions with the same symmetry in both M-theory and type IIA, with $\beta = \tfrac65$ for the SO(7) solutions and $\beta = \tfrac32$ for the $\textrm{SO}(6) \sim \textrm{SU}(4)$ solutions. The IIA G$_2$ solution happens to have $\beta = \tfrac32$ as well. Finally, ${\cal C}_2$ is the eigenvalue, normalised as indicated in footnote \ref{fn:NormCas}, of the quadratic Casimir operator in the $[p,q,r]$ representation for SO(7) or $\textrm{SU}(4)$,
{\setlength\arraycolsep{2pt}
\begin{eqnarray} \label{eq:SO(7)Casimir}
\textrm{SO}(7)  & : & {\cal C}_2 (p,q,r) = \tfrac18 \left[  4 p (p+5) + 8 q (q+4)  + 3 r (r+6) + 8 pq  + 4  pr + 8qr  \right] \;  ,   \\[5pt]
\label{eq:SU(4)Casimir}
\textrm{SU}(4)   & : &  {\cal C}_2 (p,q,r) = \tfrac18 \left[  3 p (p+4) + 4 q (q+4)  + 3 r (r+4) + 4 pq  +2 pr + 4 qr \right]  \; ,   \hspace{10pt} 
\end{eqnarray}
(or in the representation $[q,p,r]$ for SO(6), with $ {\cal C}^{\textrm{SU}(4)}_2 (p,q,r) = {\cal C}^{\textrm{SO}(6)}_2 (q,p,r)$) for the solutions with those residual symmetry groups. For the the G$_2$ solution of type IIA, ${\cal C}_2$ is the $[p,q]$ quadratic Casimir eigenvalue (\ref{eq:G2CasimirEigenv}).

Although we have not computed the KK scalar spectra for the SO(7) and $\textrm{SO}(6) \sim \textrm{SU}(4)$ solutions, and to do so is beyond the scope of this paper, it is natural to assume that these will follow the rigid pattern shown by table \ref{tab:CompleteN=0Spectra} and equation (\ref{eq:M2L2non-susy}), as the KK scalar spectrum \cite{Guarino:2020flh} of the G$_2$ solution does. Assuming that the KK scalar masses for the other solutions also take on the form (\ref{eq:M2L2non-susy}), only by choosing $\gamma_0 = 6$ and letting $\beta$ follow the pattern above, are the known spectra at KK level $n=0$ \cite{Bobev:2010ib,DallAgata:2011aa} reproduced. Level $n=0$ does not fix the coefficient $\alpha$, but it is natural to assume that this coefficient will follow the same pattern as fields of all other spins. Following this logic, one arrives at the KK scalar mass formulae marked with $^{(*)}$ in table \ref{tab:CompleteN=0Spectra}. For the type IIA SO(7) solution, the proposed mass formula reproduces the G$_2$-singlet masses at KK level $n=2$ given in table \ref{tab:G2singletSpectra} of section \ref{sec:G2SingIIA} below. 

Except for this minor caveat on the KK scalar sector, table \ref{tab:CompleteN=0Spectra} thus gives the masses in the complete KK spectrum for all the non-supersymmetric AdS$_4$ solutions under consideration in this section. Together with the table, the complete spectra are characterised by the representation content under the relevant residual symmetry group $G$, obtained as described above. For example, the mass formulae given in table \ref{tab:CompleteN=0Spectra} for the $D=11$ SO$(7)_v$ and SO$(7)_c$ solutions are identical. But the KK spectra of these two solutions are not the same: they differ in their SO(7)-representation content. For the $\cN=0$ G$_2$ solution, the G$_2$ content of the KK spectrum has been summarised up to KK level $n=3$ in tables \ref{tab:statesatlevel0G2IIA}--\ref{tab:statesatlevel3G2IIA} below. In these tables, 
each cell lists the states with a certain spin $s_0$ and the number $m$ of them (in the format $( s = s_0 ) \times m)$, in an allowed representation $[p,q]$ of G$_2$. The corresponding masses follow from table \ref{tab:CompleteN=0Spectra}. For example, table \ref{tab:statesatlevel2G2IIA} indicates the existence of one vector, two spin-$1/2$ fermions and one scalar in the $\bm{189}$ of G$_2$, with masses $14$, $\frac{35}{2}$ and $14$, respectively, at KK level 2.

\vspace{20pt}

\begin{table}[H]
\begin{center}
{\footnotesize
\begin{tabular}{|p{30mm}|p{30mm}|} 					\hline
$[0,0]$ 				& 	$[0,1]$ 							\\[1pt]
$(s=2) \times 1$			& $(s=1)\times 1$  	\\
$(s=3/2) \times 1$ 		& $(s=1/2)\times 1$  		\\
$(s=1/2) \times 1$ & \\
 $(s=0) \times 2$ &                                      \\ [5pt] \hline
$[1,0]$ 				 			\\[1pt]
 $(s=3/2) \times 1$				\\
$(s=1) \times 2$\\
$(s=1/2) \times 2 $     \\[5pt] \cline{1-1}
$[2,0]$ 			\\[1pt]
$(s=1/2) \times 1$				\\
$(s=0) \times 2$	\\[5pt] \cline{1-1}
\end{tabular}
}
\caption{\footnotesize{States at KK level $n=0$ for the $\cN=0$ G$_2$-invariant AdS$_4$ solution of type IIA.
 }\normalsize}
\label{tab:statesatlevel0G2IIA}
\end{center}
\end{table}

\vspace{15pt}

\begin{table}[H]
\begin{center}
{\footnotesize
\begin{tabular}{|p{30mm}|p{30mm}|} 					\hline
$[0,0]$ 							& 	$[0,1]$ 							\\[1pt]
 $(s=3/2) \times 1$ 				&  $(s=3/2) \times 1$  	\\
 $ (s=1) \times 2$								&  $(s=1) \times 2$	\\
 $(s=1/2) \times 1$ & $(s=1/2) \times 2$     \\
        & $(s=0) \times 2$       \\ [5pt] \hline
$[1,0]$ 							& 	$[1,1]$ 							\\[1pt]
 $(s=2) \times 1$ 			&  $(s=1) \times 1$  	\\
 $(s=3/2) \times 2$	& 		 $(s=1/2) \times 2$ 		\\
 $(s=1) \times 2$		& 		  $(s=0) \times 1$				\\[1pt]
 $(s=1/2) \times 3$&   \\
 $(s=0) \times 3$  &   \\[5pt] \hline
$[2,0]$ 				 			\\[1pt]
 $(s=3/2) \times 1$\\[1pt]
 $(s=1) \times 3$				\\
 $(s=1/2) \times 3$\\
 $(s=0) \times 1$ \\[5pt] \cline{1-1}
$[3,0]$ 			\\[1pt]
 $(s=1/2) \times 1$ 	\\
 $(s=0) \times 2$ \\[5pt] \cline{1-1}
\end{tabular}
}
\caption{\footnotesize{States at KK level $n=1$ for the $\cN=0$ G$_2$-invariant AdS$_4$ solution of type IIA.}\normalsize}
\label{tab:statesatlevel1G2IIA}
\end{center}
\end{table}

%%%%%%%%%%%

\begin{table}[H]
\begin{center}
{\footnotesize
\begin{tabular}{|p{35mm}|p{35mm}|p{35mm}|} 					\hline
$[0,0]$ 										& 	$[0,1]$ 									& 	$[0,2]$ 			\\[1pt]
 $(s=1/2) \times 1$			& $(s=3/2) \times 1$			&  $(s=1/2) \times 1  $ 	\\
 $(s=0) \times 2$					&  $(s=1) \times 3$ & 	 $(s=0) \times 2$						\\
& $(s=1/2) \times 3$ &\\
& $(s=0) \times 1$ & \\[5pt] \hline
$[1,0]$ 									& 	$[1,1]$ 			\\[1pt]
 $(s=3/2) \times 1$		&  $(s=3/2) \times 1$ 			\\[5pt] 
 $(s=1) \times 3$ 		& $(s=1) \times 3$ \\	
 $(s=1/2) \times 3$ &  $(s=1/2) \times 4$\\
 $(s=0) \times 1$  &   $(s=0) \times 3$		\\[5pt] \cline{1-2}
$[2,0]$ 									& 	$[2,1]$ 			\\[1pt]
 $(s=2) \times 1$	&  $(s=1) \times 1$ 			\\[5pt] 
 $(s=3/2) \times 2$		& 	 $(s=1/2) \times 2$	\\[5pt] 
 $(s=1) \times 2$ &  $(s=0) \times 1$	\\
 $(s=1/2) \times 4$& \\
 $(s=0) \times 5$ & \\[5pt] \cline{1-2}
$[3,0]$ 			\\[1pt]
 $(s=3/2) \times 1$ \\
 $(s=1) \times 3$	\\
 $(s=1/2) \times 3$\\
 $(s=0) \times 1$ \\[5pt] \cline{1-1}
$[4,0]$ 			\\[1pt]
 $(s=1/2) \times 1$ 	\\
 $(s=0) \times 2$  \\[5pt] \cline{1-1}
\end{tabular}
}
\caption{\footnotesize{States at KK level $n=2$ for the $\cN=0$ G$_2$-invariant AdS$_4$ solution of type IIA.}\normalsize}
\label{tab:statesatlevel2G2IIA}
\end{center}
\end{table}

%%%%%%%%%%%%

\begin{table}[H]
\begin{center}
{\footnotesize
\begin{tabular}{p{35mm}|p{35mm}|p{35mm}|}

			\cline{2-3} %\hline 
 											& 	$[0,1]$ 									& 	$[0,2]$ 			\\[1pt]
 											& $(s=1) \times 1$					& $(s=1) \times 1$	\\
 											&$(s=1/2) \times 2$  & 			$(s=1/2) \times 2$				\\
 &$(s=0) \times 1$  &$(s=0) \times 1$\\[5pt] \hline
\multicolumn{1}{ |l|  }{$[1,0]$	}									& 	$[1,1]$ 									& 	$[1,2]$		\\[1pt]
\multicolumn{1}{ |l| }{$(s=1/2) \times 1$} 			& $(s=3/2) \times 1$			& $(s=1/2) \times 1$ 	\\
\multicolumn{1}{ |l| }{$(s=0) \times 2$}			& $(s=1) \times 3$ & $(s=0) \times 2$				\\
\multicolumn{1}{ |l| }{}										&	$(s=1/2) \times 4$ & 							\\
\multicolumn{1}{ |l| }{}										&	$(s=0) \times 3$ &   \\[5pt] \hline
\multicolumn{1}{ |l|  }{$[2,0]$} 									& 	$[2,1]$ 			\\[1pt]
\multicolumn{1}{ |l| }{$(s=3/2) \times 1$} 		& $(s=3/2) \times 1$ 			\\[5pt] 
\multicolumn{1}{ |l|  }{$(s=1) \times 3$} 		&$(s=1) \times 3$ 	\\[5pt] 
\multicolumn{1}{ |l|  }{$(s=1/2) \times 3$}         & $(s=1/2) \times 4$ \\
\multicolumn{1}{ |l|  }{$(s=0) \times 1$}          & $(s=0) \times 3$      \\[5pt] \cline{1-2}
\multicolumn{1}{ |l|  }{$[3,0]$} 									& 	$[3,1]$ 			\\[1pt]
\multicolumn{1}{ |l| }{$(s=2) \times 1$} 					& $(s=1) \times 1$ 			\\
\multicolumn{1}{ |l| }{$(s=3/2) \times 2$} & $(s=1/2) \times 2$\\
\multicolumn{1}{ |l| }{$(s=1) \times 2$} & $(s=0) \times 1$\\
\multicolumn{1}{ |l| }{$(s=1/2) \times 4$} &\\
\multicolumn{1}{ |l| }{$(s=0) \times 5$} & \\[5pt]  \cline{1-2}
\multicolumn{1}{ |l|  }{$[4,0]$} 			\\[1pt]
\multicolumn{1}{ |l|  }{$(s=3/2) \times 1$} \\
\multicolumn{1}{ |l|  }{$(s=1) \times 3$}	\\
\multicolumn{1}{ |l|  }{$(s=1/2) \times 3$} \\
\multicolumn{1}{ |l|  }{$(s=0) \times 1$}  \\[5pt] \cline{1-1}
\multicolumn{1}{ |l|  }{$[5,0]$} 			\\[1pt]
\multicolumn{1}{ |l|  }{$(s=1/2) \times 1$}	\\
\multicolumn{1}{ |l|  }{$(s=0) \times 2$}	\\[5pt] \cline{1-1}
\end{tabular}
}
\caption{\footnotesize{States at KK level $n=3$ for the $\cN=0$ G$_2$-invariant AdS$_4$ solution of type IIA.}\normalsize}
\label{tab:statesatlevel3G2IIA}
\end{center}
\end{table}

\newpage

%%%%%%%%%%%%%%%%%%%%%%%
%%%%%%%%%%%%%%%%%%%%%%%

\section{Discussion} \label{sec:Discussion}

%%%%%%%%%%%%%%%%%%%%%%%
%%%%%%%%%%%%%%%%%%%%%%%

We have derived from ExFT the KK fermionic mass matrices for a class of AdS solutions of string and M-theory that uplift on spheres from maximal gauged supergravity. We have focused on E$_{7(7)}$ ExFT, but similar mass matrices can be derived for other instances of ExFT with other duality groups. We have also used these mass matrices to obtain the spectrum of KK fermions about some concrete AdS$_4$ solutions of M-theory and massive type IIA supergravity. These results, together with previously known sectors of the bosonic spectra, have allowed us to give the complete spectrum for some $\cN=1$ and some non-supersymmetric solutions in this class.

%%%%%%%%%%%%%%%%%%%%%%%

\subsection{A more general pattern for the conformal dimensions} \label{sec:GenPat}

%%%%%%%%%%%%%%%%%%%%%%%

A generic formula, (\ref{eq:E0generic}), exists for the conformal dimensions of the OSp$(4|1)$ supermultiplets present in the KK spectra of the $\cN=1$ AdS$_4$ solutions of M-theory and type IIA that we have covered in this work. The expression (\ref{eq:E0generic}) can be further generalised to account for all the spectra known so far of supersymmetric $D=11$ and type IIA AdS$_4$ solutions that uplift from the SO(8) or ISO(7) maximal supergravities. Consider an AdS$_4$ solution in this class preserving $\cN$ supersymmetries, invariant under a residual symmetry group $G \subset \textrm{SO}(8)$ in $D=11$ or $G \subset \textrm{SO}(7)$ in type IIA. We find that the dimension $E_0$ of an OSp$(4|\cN)$ supermultiplet\footnote{The dimension of a supermultiplet is defined to be the dimension of its superconformal primary.} in the KK spectrum of this solution, with superconformal primary of spin $s_0$ and arising at KK level $n$, turns out to be given generically by 
\begin{equation} \label{eq:E0genericN}
 \quad E_0 =  s_0^\2 -\tfrac12 + \sqrt{ \tfrac{9}{4} + s_0^\2 (s_0^\2 +1) -s_0(s_0+1) + \alpha \, n (n + d-1) + {\cal Q}_2} \; .
\end{equation}
Here, $n (n + d-1)$ is the eigenvalue of the scalar Laplacian on $S^d$, with $d=7$ for $D=11$ and $d=6$ for type IIA; $\alpha$ is a solution-dependent constant; ${\cal Q}_2$ is a solution-dependent homogeneous quadratic polynomial in the integer Dynkin labels of $G$; and 
\begin{eqnarray}
\label{s02}
s_0^\2 = 
\left\{
\begin{array}{lll}
\tfrac12 \, (4- \cN) & ,  & \quad \textrm{ if $ 1 \leq \cN \leq 4 $ }  \\[4pt]
0  & , & \quad  \textrm{ if $ 4 \leq \cN \leq 8 $ } 
\end{array}
\right.
\end{eqnarray} 
is the spin of the superconformal primary of any of the Osp$(4|\cN)$ supermultiplets containing a graviton as its highest spin state. For $\cN=1$ supersymmetry, the massless (MGRAV, in the notation of table \ref{tab:OSp(4|1) supermultiplets}) or generic (GRAV) graviton supermultiplets have $s_0^\2 = \tfrac32$, and (\ref{eq:E0generic}) is indeed of the form (\ref{eq:E0genericN}) with ${\cal Q}_2 \equiv - \beta \, {\cal C}_2 (p,q) $, for the particular values of the constant $\beta$ specified in the text and the relevant quadratic Casimir eigenvalues ${\cal C}_2 (p,q) $ in (\ref{eq:G2CasimirEigenv}) or (\ref{eq:SU3CasimirEigenv}). 

Formula (\ref{eq:E0genericN}) also describes the spectrum for all the $\cN \geq 2$ solutions in the class we are considering. Specifically,  a generic formula that agrees with (\ref{eq:E0genericN}) can be written for the dimensions of the OSp$(4|2)$ supermultiplets present in the KK spectrum of the $\cN=2$ $\textrm{SU}(3) \times \textrm{U}(1)$-invariant AdS$_4$ solutions of M-theory \cite{Corrado:2001nv} and type IIA \cite{Guarino:2015jca}. From the spectral results for these solutions \cite{Malek:2020yue,Varela:2020wty}, it follows that the dimension $E_0$ of an OSp$(4|2)$ supermultiplet with conformal primary spin $s_0$, present in the spectrum at KK level $n$ with $\textrm{SU}(3) \times \textrm{U}(1)$ charges $[p,q]_{y_0}$ is
\begin{equation} \label{eq:E0genericN=2}
\cN=2 \; : \quad E_0 = \tfrac12 + \sqrt{ \tfrac{17}{4} -s_0(s_0+1) + \alpha \, n (n + d-1) - \tfrac43 \, {\cal C}_2 (p,q) + \tfrac12 \, y_0^2} \; .
\end{equation}
Here, $ {\cal C}_2 (p,q)$ is again the SU(3) Casimir eigenvalue (\ref{eq:SU3CasimirEigenv}), and now $d=7$, $\alpha= \tfrac12$ for the  $D=11$ solution \cite{Corrado:2001nv} and $d=6$, $\alpha=\tfrac23$ for the type IIA one \cite{Guarino:2015jca}. With these definitions, (\ref{eq:E0genericN=2}) agrees with the expressions provided in \cite{Malek:2020yue,Varela:2020wty} for the various OSp$(4|2)$ supermultiplets, including the hypermultiplets. In order to compare (\ref{eq:E0genericN=2}) with the expressions provided in those references recall that $\cN=2$ (massless, short and long) graviton, (short and long) gravitino, (massless, short and long) vector multiplets, and hypermultiplets respectively have $s_0 \equiv s_0^\2 = 1$, $s_0 =\tfrac12$, $s_0=0$ and $s_0=0$. Since (massless, short or long) $\cN=2$ graviton multiplets have $s_0^\2 = 1$, (\ref{eq:E0genericN=2}) is also of the form (\ref{eq:E0genericN}) with ${\cal Q}_2 \equiv - \tfrac43 \, {\cal C}_2 (p,q) + \tfrac12 \, y_0^2$. 

A similar observation holds for the $\cN=3$ AdS$_4$ solution of type IIA \cite{Gallerati:2014xra,Pang:2015vna,DeLuca:2018buk} with $ \textrm{SO}(3)_{\cal R} \times \textrm{SO}(3)_{\cal F}$ invariance. It follows from \cite{Varela:2020wty} that an OSp$(4|3)$ supermultiplet with conformal primary spin $s_0$, present at KK level $n$ with $ \textrm{SO}(3)_{\cal R} \times \textrm{SO}(3)_{\cal F}$ quantum numbers $(j,h)$ has conformal dimension
\begin{equation} \label{eq:E0genericN=3}
\cN=3 \; : \quad E_0 = \sqrt{ 3 -s_0(s_0+1) + \tfrac12 \, n (n + d-1) + \tfrac12 \, j ( j +1)  - \tfrac32 \, h ( h +1) } \; .
\end{equation}
With $d=6$, this formula indeed reproduces (4.6) and (4.7) of \cite{Varela:2020wty} for the (massless, short and long) graviton and the (short and long) gravitino multiplets. These have respectively $s_0  \equiv s_0^\2 = \tfrac12$ and $s_0 = 0$, see {\it e.g.} appendix B of \cite{Varela:2020wty}. Equation (\ref{eq:E0genericN=3}) also reproduces the dimension for the (necessarily short for $\cN=3$) vector multiplets, which have $s_0 = 0$. This was given in table 8 of \cite{Varela:2020wty}. Indeed, (\ref{eq:E0genericN=3}) reduces to $E_0 = \frac12 (n+2)$ as given in that table upon using that (the unique) vector multiplet at KK level $n$ has quantum numbers $ j = h = \frac12 (n+2) $. Further, (\ref{eq:E0genericN=3}) is also of the generic form (\ref{eq:E0genericN}) with $\alpha = \tfrac12$, given that the (massless, short, or long) $\cN=3$ graviton multiplet has $s_0^\2 = \tfrac12$. In this case, ${\cal Q}_2 \equiv \tfrac12 \, j ( j +1)  - \tfrac32 \, h ( h +1)$. 

Finally, the dimension of the (unique) Osp$(4|8)$ supermultiplet present at level $n$ in the KK spectrum \cite{Englert:1983rn,Sezgin:1983ik,Biran:1983iy} of the $\cN=8$ Freund-Rubin solution \cite{Freund:1980xh} of $D=11
$ supergravity is $E_0 = \tfrac12 (n+2)$ (see {\it e.g.} table 9 of \cite{Duff:1986hr}). This may be straightforwardly rewritten as 
\begin{equation} \label{eq:E0genericN=8}
\cN=8 \; : \quad E_0 = -\tfrac12 +  \sqrt{ \tfrac94 + \tfrac14 \, n (n + d-1) } 
\end{equation}
with $d=7$. Since these Osp$(4|8)$ multiplets all have scalar, $s_0 = s_0^\2 = 0 $, superconformal primaries, (\ref{eq:E0genericN=8}) also conforms to the generic expression (\ref{eq:E0genericN}) with ${\cal Q}_2 \equiv 0 $. 

Of course, the formulae (\ref{eq:E0generic}), (\ref{eq:E0genericN}) may not necessarily extrapolate to other supersymmetric AdS$_4$ solutions of M-theory and type IIA that still uplift from the SO(8) or ISO(7) gaugings, but preserve other symmetry groups.

 \newpage

%%%%%%%%%%%%%%%%%%%%%%%

\subsection{G$_2$-singlet spectra in type IIA and consistent truncations} \label{sec:G2SingIIA}

%%%%%%%%%%%%%%%%%%%%%%%

On a different note, table \ref{tab:multipletsatlevel3G2IIA} of section \ref{sec:N=1G2SpectrumIIA} shows that there are no G$_2$-singlet supermultiplets at KK level $n=3$ for the $\cN=1$ G$_2$-invariant solution \cite{Behrndt:2004km} of type IIA. The claim is in fact stronger: the number of G$_2$ singlets in the KK spectrum of this solution is finite, there are no singlets for $n \geq 3$, and all of them appear at levels $n=0,1,2$. This can be seen by branching the SO(7) representations in table 1 of \cite{Varela:2020wty} under $\textrm{G}_2 \subset \textrm{SO}(7)$ for all $n$. The same holds for the non-supersymmetric G$_2$-invariant solution of type IIA (see section \ref{sec:N=0G2SpectrumIIA}), relative to the individual G$_2$-singlet KK states as there is obviously no supermultiplet structure in that case. For the non-supersymmetric SO(7)-invariant solution of type IIA, something similar happens: its complete KK spectrum comes in an infinite number of SO(7) representations, but the number of singlets under the branching $\textrm{SO}(7) \supset \textrm{G}_2$ is also finite. The complete spectrum of G$_2$-singlet states for each of these three AdS$_4$ solutions of type IIA is summarised in table \ref{tab:G2singletSpectra}. For the $\cN=1$ G$_2$ solution, the states combine  into the OSp$(4|1)$ supermultiplets indicated in the table, as follows from tables \ref{tab:multipletsatlevel0G2IIA}--\ref{tab:multipletsatlevel2G2IIA}.

This feature of the KK spectra for these three AdS$_4$ solutions was expected on the following grounds. Massive type IIA supergravity admits a fully non-linear consistent truncation on $S^6$ down to a certain $D=4$ $\cN=2$ gauged supergravity coupled to a vector multiplet and a hypermultiplet \cite{Cassani:2009ck}. This truncation is obtained by expanding the type IIA fluxes along the forms that define the canonical, homogeneous nearly-K\"ahler structure on $S^6$ with $D=4$ field coefficients, and is in fact valid for any nearly-K\"ahler six-dimensional manifold \cite{KashaniPoor:2007tr}. This $D=4$ $\cN=2$ theory \cite{Cassani:2009ck} is not contained in $D=4$ $\cN=8$ ISO(7) supergravity. Rather, these two theories overlap \cite{Guarino:2015vca} in the G$_2$-invariant sector \cite{Guarino:2015qaa} of the latter. The $\cN=8$ supergravity captures the modes at KK level $n=0$ in the compactification of massive IIA on $S^6$ and reconstructs their full non-linear interactions. It was argued in \cite{Guarino:2015vca} that the $\cN=2$ theory should do likewise for the G$_2$-singlet states up the KK towers around any of its three vacua (thereby identified with the three vacua of the $\cN=8$ ISO(7) theory that appear in table \ref{tab:G2singletSpectra}).

 \begin{table}[]

%\centering

\resizebox{\textwidth}{!}{

\begin{tabular}{|l|c|cccccccccccc|l|}
\hline
Sol.                & $n$                  & $s=2$     & $s=\tfrac32$     & $s=\tfrac32$     & $s=1$     & $s=1$     & $s=\tfrac12$     & $s=\tfrac12$     & $s=\tfrac12$     & $s=0$     & $s=0$     & $s=0$     & $s=0$               &         OSp$(4|1)$ supermultiplet             \\ \hline \hline
\multirow{4.0}{*}{$\cN=1$, G$_2$} & $0$    & $ 0 $        & $1$       & $$   & $$    & $$   & $$   & $\sqrt{6}$ & $$   & $4-\sqrt{6}$ & $4+\sqrt{6}$         & $$    & $$           & MGRAV $ \big( \tfrac52 \big)$   , CHIRAL $  \left( 1+ \sqrt{6}  \right) $        \\[4pt]  
\cline{2-15} %\cline{7} 
  & 1 &   $$ &   $$  & $3$ & $12$ & $6$ & $3$  & $$ & $$   & $$ & $$         & $$    & $$       &  GINO $ ( 4 )$   \\[4pt]  
\cline{2-15} %\cline{7} 
& $ 2$ & $ $        & $$       & $$   & $$    & $$   & $$   & $$ & $\sqrt{\tfrac{53}{3}}$   & $$ & $$         & $ \tfrac{47}{3} - \sqrt{\tfrac{53}{3}} $    & $\tfrac{47}{3} + \sqrt{\tfrac{53}{3}}$     &  CHIRAL $ \left( 1+ \sqrt{\tfrac{53}{3}}  \right) $    \\[4pt]  \hline \hline
\multirow{4}{*}{$\cN=0$, SO$(7)$} & $0$    & $0$     &  $\tfrac{3}{2} \sqrt{\tfrac{3}{5}}$  & $$ & $$  & $$  & $$     & $- $ & $$ & $6$   & $-\tfrac{6}{5}$          & $$    & $$      & $-$ \\[4pt]  
\cline{2-15} %\cline{7} 
& $ 1$ &   $$ &   $$  & $\tfrac{7}{2} \sqrt{\tfrac{3}{5}}$  & $12$ & $\tfrac{24}{5}$ & $\tfrac32 \sqrt{\tfrac{23}{5}}$  & $$ & $$   & $$ & $$         & $$    & $$       & $-$
 \\[4pt]  
\cline{2-15} %\cline{7} 
& $ 2$ & $ $        & $$       & $$   & $$    & $$   & $$   & $$ & $\tfrac12 \sqrt{\tfrac{367}{5}} $   & $$ & $$         & $ 20 $    & $\tfrac{64}{5}$     & $-$
\\[4pt]  \hline \hline
\multirow{4}{*}{$\cN=0$, G$_2$} & $0$    & $0$       & $\tfrac{3}{\sqrt{2}}$   & $$   & $$    & $$   & $$   & $-$ & $$   & $6$ & $6$         & $$    & $$    & $-$       \\[4pt]  
\cline{2-15} %\cline{7} 
& $ 1$ & $$     & $$      & $\sqrt{\tfrac{21}{2}}$ & $12$ & $12$ & $\tfrac{3\sqrt{6}}{2}$  & $$ & $$   & $$ & $$         & $$    & $$    & $-$
 \\[4pt]  
\cline{2-15} %\cline{7} 
& $ 2$  & $ $        & $$       & $$   & $$    & $$   & $$   & $$ & $\sqrt{\tfrac{43}{2}}$   & $$ & $$         & $ 20 $    & $20$  & $-$
\\[4pt]  \hline 
\end{tabular}
}

\caption{\footnotesize{All G$_2$-singlet states in the complete KK spectra of the AdS$_4$ solutions of type IIA indicated on the left-most column. The KK level $n$ at which each state arises, its spin $0 \leq s \leq 2$, and the OSp$(4|1)$ supermultiplet with the given dimension to which it belongs, if appropriate, are indicated. Each entry gives the mass (squared, $m^2L^2$, for the bosons and linear, $|mL|$, for the fermions) for each state in units of the AdS$_4$ radius $L$. These G$_2$-singlet spectra are also the spectra of the $D=4$ $\cN=2$ theory of \cite{Cassani:2009ck} around each of its three AdS vacua.
}\normalsize}
\label{tab:G2singletSpectra}
\end{table}

For this picture to hold, the number of G$_2$-singlets in the KK spectra about any of these solutions should be finite, precisely as we find. Further, the scalar spectrum for these solutions computed here and in \cite{Guarino:2020flh} from ExFT, precisely matches the spectrum computed within the $\cN=2$ theory of \cite{Cassani:2009ck} in table 1 of \cite{Guarino:2015vca}. Table \ref{tab:G2singletSpectra} thus gives the full spectrum, including the fermions, of the $\cN=2$ theory of \cite{Cassani:2009ck} around each if its three vacua, further identifying the KK level at which each mode arises. 

The situation in $D=11$ is similar, not with respect to G$_2$, but with respect to SU$(4)_c$ or SU$(4)_s$. From section \ref{sec:N=1G2SpectrumIIA}, the KK spectrum of the $\cN=1$ G$_2$ solution of $D=11$ supergravity can be seen to include an infinite number of G$_2$-singlet states, unlike its type IIA counterpart. In $D=11$, it is instead the KK spectra of the $\cN=8$ SO(8), the $\cN=0$ SO(7)$_c$, and the $\cN=0$ SU(4)$_c$ solutions that contain a finite number of SU(4)$_c$--singlet states. These states are retained in a $D=4$ $\cN=2$ consistent truncation of $D=11$ supergravity on $S^7$ (or any other Sasaki-Einstein manifold) \cite{Gauntlett:2009zw,Gauntlett:2009bh}, which is different from the $\cN=8$ truncation \cite{deWit:1986iy} to the SO(8)-gauged supergravity \cite{deWit:1982ig}, but overlaps with it \cite{Bobev:2010ib} in the SU$(4)_c$-invariant sector of the latter. Thus, the truncation of \cite{Gauntlett:2009zw,Gauntlett:2009bh} also retains higher KK modes in the $D=11$ case. The KK spectrum of the $\cN=8$ SO(8) point, and only of this point,  also contains a finite number of modes when branched under $\textrm{SU}(4)_s$. These are simply the states contained in the $\cN=2$ supergravity multiplet. The resulting truncation to minimal $D=4$ $\cN=2$ gauged supergravity \cite{Larios:2019kbw} agrees with that discussed more generally in \cite{Gauntlett:2007ma}.

Other consistent truncations of $D=11$ \cite{Cremmer:1978km} and type IIB \cite{Schwarz:1983qr} supergravities down to lower-dimensional gauged supergravities are known that, similarly to \cite{Cassani:2009ck,Gauntlett:2009zw,Gauntlett:2009bh}, keep singlet modes up the corresponding KK towers and reconstruct their non-linear interactions \cite{Cassani:2011fu,Cassani:2012pj,Cassani:2010uw,Gauntlett:2010vu,Liu:2010sa,Donos:2010ax}. The systematics of this type of ``massive mode truncations" has been recently elucidated \cite{Cassani:2019vcl,Malek:2019ucd} using duality-covariant techniques \cite{Pacheco:2008ps,Berman:2010is,Hohm:2013pua,Hohm:2013vpa,Hohm:2013uia,Ciceri:2016dmd,Cassani:2016ncu}, yet another demonstration of the power of these methods for supergravity applications.

%%%%%
%%%%%

\section*{Acknowledgements}

%%%%%
%%%%%

We would like to thank Gabriel Larios for useful discussions. MC is supported by a La Caixa Foundation (ID 100010434) predoctoral fellowship LCF/ BQ/DI19/11730027. OV is supported by the NSF grant PHY-2014163. MC and OV are partially sup\-por\-ted by grants SEV-2016-0597 and PGC2018-095976-B-C21 from MCIU/ 
   AEI/FEDER, UE.

\newpage

%%%%%
%%%%%

\appendix

%%%%%
%%%%%

\addtocontents{toc}{\setcounter{tocdepth}{1}}

\section{KK fermion spectra of selected AdS$_4$ solutions} \label{sec:SpecificSpectra}

Our conventions for $D=11$ \cite{Cremmer:1978km} and $D=4$ $\cN=8$ SO(8) \cite{deWit:1982ig} supergravity are those of \cite{Varela:2015ywx}. For massive type IIA supergravity \cite{Romans:1985tz} and $D=4$ $\cN=8$ ISO(7) supergravity we follow \cite{Guarino:2015vca} and \cite{Guarino:2015qaa}, respectively. For generic  conventions on $D=4$ $\cN=8$ gauged supergravity \cite{deWit:2007mt} we follow \cite{Guarino:2015vca}. See appendix A of \cite{Varela:2020wty} for the explicit expressions of the embedding tensors that appear in the fermionic mass matrices given in section \ref{sec:KKFermionMassMat} of the main text. See also the appendix of \cite{Varela:2020wty} for explicit expressions of the SO(8) or SO(7) generators $({\cal T}_{\underline{N}})_\Lambda{}^\Sigma$ in our conventions.

We have employed our mass matrices to compute the first few levels of the KK fermion spectrum of some AdS$_4$ solutions of $D=11$ and massive IIA supergravity that uplift from critical points of SO(8) or ISO(7) $D=4$ $\cN=8$ supergravities. For concreteness, we have focused on solutions that preserve at least SU(3) symmetry. The $D=4$ critical points of SO(8) supergravity in this sector were classified in \cite{Warner:1983vz}, and their $D=11$ uplift is known \cite{Freund:1980xh,Corrado:2001nv,deWit:1984nz,Godazgar:2013nma,deWit:1984va,Englert:1982vs,Pope:1984bd}. The entire KK spectrum of the $\cN=8$ SO(8)-invariant solution \cite{Freund:1980xh} has long been known \cite{Englert:1983rn,Sezgin:1983ik,Biran:1983iy}, and our results reproduce their KK fermion spectrum. The complete spectrum of the 
$\cN=2$ $\textrm{SU}(3) \times \textrm{U}(1)$ solution \cite{Warner:1983vz,Corrado:2001nv} is now also known \cite{Klebanov:2008vq,Malek:2020yue,Varela:2020wty}. We again reproduce the KK gravitino and spin-$1/2$ spectra for this solution. For all other solutions, only the spectrum of KK gravitons \cite{Klebanov:2009kp} and vectors \cite{Dimmitt:2019qla} are known, and our fermionic spectra are new. 

The results are summarised for the KK gravitini at levels $n=0,1,2$ in table \ref{tab:KKGravitiniSO8}. The table lists the KK gravitini linear masses without sign, $|ML|$, normalised to the corresponding AdS$_4$ radius, $L = \sqrt{-6/V}$. Here, $V<0$ is the cosmological constant of each AdS$_4$ critical point. The eigenvalues in the table appear as $ |M L|^{(p)}$, where $p$ is a positive integer that denotes the multiplicity. Recall that the scaling dimension of a gravitino \cite{Volovich:1998tj} or a spin-$1/2$ fermion \cite{Henningson:1998cd} on AdS$_4$ of mass $ | M L|$ is given by 
\begin{equation} \label{eq:DeltaFermions}
\Delta = \tfrac32 +   |M L| \; .
\end{equation}
This formula has been used throughout to convert the KK fermion mass eigenvalues to the conformal dimensions reported in the main text. We note that, for the $\cN=2$ AdS$_4$ solution \cite{Warner:1983vz,Corrado:2001nv}, one needs to take (\ref{eq:DeltaFermions}) without absolute value and with negative mass, $ML <0$ for some spin-$1/2$ states, in order to reproduce the spectrum that can be deduced from the bosonic calculation of \cite{Malek:2020yue,Varela:2020wty}. For completeness, recall that for gravitons and scalars the relation between dimension and mass is
\begin{equation}
\Delta (\Delta-3) = M^2 L^2 \; ,
\end{equation}
while for vectors one has
\begin{equation}
(\Delta - 1)  (\Delta-2) = M^2 L^2 \; .
\end{equation}
%

%\begin{landscape}

 \begin{table}%[H]
\centering

\resizebox{\textwidth}{!}{

%\tiny{

\begin{tabular}{|l|l|l|}
\hline
Sol.                & $n$                  & $\lvert M L\rvert$                                     \\ \hline \hline
\multirow{5}{*}{$\cN=0$, G$_2$} & $0$    & $\frac{3}{\sqrt{2}}^{(7)}$,  $\sqrt{\frac{3}{2}}^{(14)}$ ,  $\sqrt{\frac{1}{2}}^{(27)}$                             \\[4pt]  
\cline{2-3} %\cline{7} 
  & 1 &    $\sqrt{\frac{27}{2}}^{(22)}$ ,   $\sqrt{\frac{15}{2}}^{(28)}$ ,    $\sqrt{\frac{13}{2}}^{(81)}$ ,   $\sqrt{3}^{(128)}$ , $\sqrt{\frac{3}{2}}^{(77)}$   \\[4pt]  
\cline{2-3} %\cline{7} 
& 2 &        $\sqrt{\frac{43}{2}}^{(1)}$  , $\sqrt{\frac{37}{2}}^{(21)}$ , $\sqrt{\frac{31}{2}}^{(42)}$,  $\sqrt{\frac{29}{2}}^{(108)}$ ,  $\sqrt{11}^{(256)}$,   $\sqrt{\frac{19}{2}}^{(231)}$    , $\sqrt{\frac{13}{2}}^{(77)}$   ,    $\sqrt{\frac{7}{2}}^{(182)}$  , $\sqrt{\frac{11}{2}}^{(378)}$   
\\[4pt]  \hline 
\end{tabular}

%}

}\normalsize

\caption{\footnotesize{The KK spin-$1/2$ spectra up to KK level $n=2$
about the massive type IIA AdS$_4$ solution that uplifts on $S^6$ from the $\cN=0$ G$_2$-invariant vacuum of $D=4$ $\cN=8$ ISO(7) dyonically-gauged supergravity.}\normalsize}
\label{tab:KKSpin1/2G2N=0}
\end{table}

%\end{landscape}

For the massive type IIA solutions we have again focused on the solutions that preserve the SU(3) subgroup of SO(7). These solutions have been classified in $D=4$ $\cN=8$ ISO(7) supergravity \cite{Guarino:2015qaa}, and uplifted to ten dimensions \cite{Guarino:2015jca,Varela:2015uca,Pang:2015vna,DeLuca:2018buk}. The bosonic spectrum for all these solutions is known at KK level $n=0$ \cite{Guarino:2015qaa}, and at all KK levels $n \geq 0$ for the gravitons \cite{Pang:2017omp}, vectors \cite{Varela:2020wty} and, for the $\cN=2$ $\textrm{SU}(3) \times \textrm{U}(1)$ \cite{Varela:2020wty} and $\cN=0$ G$_2$ \cite{Guarino:2020flh} solutions in this class, also for the scalars. The complete $\cN=2$ spectrum for the $\cN=2$ $\textrm{SU}(3) \times \textrm{U}(1)$ solution is in fact known \cite{Varela:2020wty}, and we reproduce the fermionic sector. For all other solutions, the fermion spectra are new. As a sample of our results for the spin-$1/2$ spectrum, we include table \ref{tab:KKSpin1/2G2N=0} with the eigenvalues up to KK level $n=2$ for the $\cN=0$ G$_2$ solution in the ISO(7) gauging. The results for the KK gravitino masses up to level $n=2$ are summarised in table \ref{tab:KKGravitiniISO7}. The format and conventions are the same as table \ref{tab:KKGravitiniSO8}.

%%%%%%%%%%%%%%%%%%%%%%%%%%%%%%%%

%\newpage

\begin{landscape}
\begin{table}%[H]
\centering
%\resizebox{\textwidth}{!}{

\footnotesize{

\begin{tabular}{|l|l|l|}
\hline
Sol.                & $n$                  & $ \lvert M L \rvert$                                     \\ \hline \hline
\multirow{4}{*}{$\cN=8$, SO(8)} & $0$    & $1^{(8)}$                                \\[4pt]  
\cline{2-3} %\cline{7} 
  & 1 & $\tfrac{5}{2}^{(8)}$ , $\tfrac{3}{4}^{(56)}$          \\[4pt]  
\cline{2-3} %\cline{7} 
& $ 2$ & $3^{(56)}$, $2^{(224)}$                              \\[4pt]  \hline \hline
\multirow{6}{*}{$\cN=2$, U(3)} & $0$    &    $\tfrac{4}{3}^{(6)}$ ,   $1^{(2)}$                            \\[4pt]  
\cline{2-3} %\cline{7} 
  & 1 &   $3^{(2)}$, $\frac{1}{6}(\sqrt{145}\pm 3)^{(12)}$ ,    $\frac{7}{2}^{(6)}$ ,  $2^{(4)}$ , $\sqrt{3}^{(16)}$ , $\frac{5}{3}^{(12)}$   \\[4pt]  
\cline{2-3} %\cline{7} 
& \multirow{2.8}{*}{$2$} &     $4^{(2)}$,   $\frac{1}{2}(\sqrt{41}\pm 1)^{(2)}$  , $\frac{1}{6}(\sqrt{337}\pm 3)^{(12)}$, $\frac{1}{6}(\sqrt{313}\pm 3)^{(12)}$, $\frac{10}{3}^{(6)}$ , $\tfrac{\sqrt{88}}{3}^{(12)}$ , $3^{(20)}$, $\frac{1}{6}(\sqrt{217}\pm 3)^{(24)}$, $\sqrt{8}^{(32)}$, $\frac{8}{3}^{(12)}$,
\\[4pt] 
&  &     $\tfrac{\sqrt{40}}{3}^{(60)}$ ,  $\sqrt{2}^{(35)}$  , $1^{(1)}$        \\[4pt]  \hline \hline
\multirow{4.5}{*}{$\cN=1$, G$_2$} & $0$    & $\sqrt{\tfrac{3}{2}}^{(7)}$    , $1^{(1)}$                            \\[4pt]  
\cline{2-3} %\cline{7} 
  & 1 & $\frac{1}{4}(\sqrt{106}\pm 2)^{(1)}$ ,   $\frac{1}{4}(\sqrt{66}\pm 2)^{(7)}$   ,   $\sqrt{\tfrac{47}{8}}^{(7)} $   , $ \sqrt{\tfrac{27}{8}}^{(14)}$ ,   $\sqrt{\tfrac{61}{24}}^{(27)} $   \\[4pt]  
\cline{2-3} %\cline{7} 
& $ 2$ & $4^{(1)}$, $ \frac{1}{2}(\sqrt{39}\pm 1)^{(7)}$    , $\sqrt{\tfrac{23}{2}}^{(14)} $  , $ \frac{1}{6}(\sqrt{231}\pm3)^{(27)}$  , $3^{(29)}$,  $\frac{7}{\sqrt{6}}^{(27)}$   ,  $\tfrac{\sqrt{21}}{2}^{(64)} $ , ${2^{(77)} }$  \\[4pt]  \hline \hline
\multirow{4}{*}{$\cN=0$, SO$(7)_v$} & 0    & $\frac{1}{2}\sqrt{\frac{27}{5}}^{(8)}$          \\[4pt]  
\cline{2-3} %\cline{7} 
& $ 1$ &   $\sqrt{\frac{33}{5}}^{(16)}$ , $2\sqrt{\frac{3}{5}}^{(48)}$ 
 \\[4pt]  
\cline{2-3} %\cline{7} 
& $ 2$ &    $\frac{1}{2}\sqrt{\frac{267}{5}}^{(16)}$ , $\frac{1}{2}\sqrt{\frac{183}{5}}^{(96)}$ ,  $\frac{\sqrt{15}}{2}^{(168)}$ 
\\[4pt]  \hline \hline
\multirow{4}{*}{$\cN=0$, SO$(7)_c$} & $0$    & $\frac{1}{2}\sqrt{\frac{27}{5}}^{(8)}$             \\[4pt]  
\cline{2-3} %\cline{7} 
& $ 1$ &   $\frac{\sqrt{39}}{2}^{(1)}$ , $\frac{1}{2}\sqrt{\frac{123}{5}}^{(7)}$ , $\frac{\sqrt{15}}{2}^{(21)}$, $\frac{1}{2}\sqrt{\frac{51}{5}}^{(35)}$ 
 \\[4pt]  
\cline{2-3} %\cline{7} 
& $ 2$ &    $\frac{1}{2}\sqrt{\frac{267}{5}}^{(8)}$ , $\frac{1}{2}\sqrt{\frac{183}{5}}^{(48)}$ ,  $\frac{1}{2}\sqrt{\frac{123}{5}}^{(112)}$ ,  $\frac{1}{2}\sqrt{\frac{123}{5}}^{(112)}$
\\[4pt]  \hline \hline
\multirow{4}{*}{$\cN=0$, SU$(4)_c$} & $0$    &   $\frac{\sqrt{27}}{4}^{(8)}$             \\[4pt]  
\cline{2-3} %\cline{7} 
& $ 1$ &   $\frac{\sqrt{39}}{2}^{(2)}$ , $\sqrt{6}^{(12)}$ , $\frac{\sqrt{15}}{2}^{(30)}$, $\sqrt{3}^{(20)}$ 
 \\[4pt]  
\cline{2-3} %\cline{7} 
& $ 2$ &  $\frac{\sqrt{219}}{4}^{(16)}$, $\frac{7}{4}\sqrt{3}^{(80)}$, $\frac{3}{4}\sqrt{11}^{(144)}$, $\frac{5}{4}\sqrt{3}^{(40)}$
\\[4pt]  \hline 
\end{tabular}

%}

}\normalsize

\caption{\footnotesize{The KK gravitino spectra up to KK level $n=2$
about the $D=11$ AdS$_4$ solutions that uplift on $S^7$ from vacua of $D=4$ $\cN=8$ SO(8) gauged supergravity with at least SU(3) symmetry.}\normalsize}
\label{tab:KKGravitiniSO8}
\end{table}

\end{landscape}

%%%%%%%%%%%%%%%%%%%%%%%%%%%%%%%%

\newpage

\begin{landscape}

 \begin{table}[H]
\centering

%\resizebox{\textwidth}{!}{

\footnotesize{

\begin{tabular}{|l|l|l|}
\hline
Sol.                & $n$                  & $\lvert M L\rvert$                                     \\ \hline \hline
\multirow{6}{*}{$\cN=2$, U(3)} & $0$    & $\frac{4}{3}^{(6)}$ ,  $1^{(2)}$                             \\[4pt]  
\cline{2-3} %\cline{7} 
  & 1 &    $3^{(2)}$ ,   $\frac{8}{3}^{(6)}$ ,    $\frac{2}{3}\sqrt{13}^{(6)}$ ,   $2^{(18)}$ , $6^{(20)}$ , $\frac{2}{3}\sqrt{7}^{(12)}$ ,  $\tfrac{5}{3}^{(12)}$  \\[4pt]  
\cline{2-3} %\cline{7} 
& 2 &        $\left( \sqrt{\tfrac{139}{3}} \pm \tfrac{1}{2} \right)^{(2)}$  , $\sqrt{\frac{121}{9}}^{(12)}$ , $\frac{10}{3}^{(24)}$,  $\left( \sqrt{\tfrac{91}{12}} \pm \tfrac{1}{2} \right)^{(16)}$ ,  $\sqrt{\frac{28}{3}}^{(16)}$,   $\sqrt{\frac{76}{9}}^{(12)}$    , $\sqrt{\frac{64}{9}}^{(42)}$   ,    $\sqrt{\frac{52}{9}}^{(30)}$  , $\frac{7}{3}^{(24)}$    ,    $\sqrt{\frac{16}{3}}^{(20)}$   
\\[4pt]  \hline \hline
\multirow{4.5}{*}{$\cN=1$, G$_2$} & $0$    & $\sqrt{ \tfrac{3}{2}}^{(7)}$    , $1^{(1)}$                            \\[4pt]  
\cline{2-3} %\cline{7} 
  & 1 &   $3^{(1)}$ ,  $\frac{1}{2}\left(\sqrt{19} \pm 1  \right)^{(7)}$ ,  $2^{(14)}$ ,   $ \sqrt{\tfrac{19}{6}}^{(27)}$    \\[4pt]  
\cline{2-3} %\cline{7} 
& $ 2$ & $ \sqrt{\tfrac{79}{6}}^{(7)}$  , $\left(\sqrt{\tfrac{97}{12}}\pm \frac{1}{2}\right)^{(27)} $ , $ \sqrt{\tfrac{32}{3}}^{(14)}$, $\sqrt{\tfrac{83}{12}}^{(64)}$  , $ \sqrt{\tfrac{17}{3}}^{(77)}$         \\[4pt]  \hline \hline
\multirow{5}{*}{$\cN=1$, SU(3)} & $0$    & $2^{(1)}$ ,   $\tfrac{4}{3}^{(6)}$ ,  $1^{(1)}$                           \\[4pt]  
\cline{2-3} %\cline{7} 
  & 1 &    $\frac{1}{2}\left(\sqrt{29}\pm 1\right)^{(1)}$  , $3^{(2)}$  ,  $\frac{1}{6}\left(\sqrt{181}\pm 3\right)^{(1)}$ , $ \sqrt{\tfrac{61}{9}}^{(12)}$ ,   $2^{(16)}$ ,  $\frac{\sqrt{31}}{3}^{(12)}$       \\[4pt]  
\cline{2-3} %\cline{7} 
& 2 &     $\left( \sqrt{\frac{167}{12}}\pm \frac{1}{2} \right)^{(1)}$ , $\sqrt{\tfrac{47}{3}}^{(2)}$ , $\left( \sqrt{\frac{421}{36}}\pm \frac{1}{2} \right)^{(6)}$, $\sqrt{\tfrac{121}{9}}^{(24)}$ , $\left( \sqrt{\frac{107}{12}}\pm \frac{1}{2} \right)^{(8)}$ , $\left( \sqrt{\frac{301}{36}}\pm \frac{1}{2} \right)^{(12)}$,   $ \sqrt{\tfrac{32}{3}}^{(32)}$ , $\sqrt{\tfrac{91}{9}}^{(24)}$ , $\sqrt{\tfrac{61}{9}}^{(60)}$ , $\sqrt{\tfrac{17}{3}}^{(20)}$ 
     \\[4pt]  \hline \hline
\multirow{4}{*}{$\cN=0$, SO$(7)$} & $0$    & $ \tfrac{3}{2}\sqrt{\tfrac{3}{5}}^{(8)}$            \\[4pt]  
\cline{2-3} %\cline{7} 
& $ 1$ &   $\tfrac{7}{2}\sqrt{\tfrac{3}{5}}^{(8)}$ ,  $\tfrac{3}{2}\sqrt{\tfrac{7}{5}}^{(48)}$
 \\[4pt]  
\cline{2-3} %\cline{7} 
& $ 2$ &    $\frac{1}{2}\sqrt{\frac{233}{5}}^{(48)}$ , $\frac{\sqrt{23}}{2}^{(168)}$ 
\\[4pt]  \hline \hline
\multirow{4}{*}{$\cN=0$, SO$(6)$} & $0$    &   $\frac{\sqrt{27}}{4}^{(8)}$           \\[4pt]  
\cline{2-3} %\cline{7} 
& $ 1$ & $\frac{\sqrt{123}}{4}^{(16)}$,   $\frac{\sqrt{51}}{4}^{(40)}$ 
 \\[4pt]  
\cline{2-3} %\cline{7} 
& $ 2$ &   $\frac{\sqrt{251}}{4}^{(16)}$ , $\frac{\sqrt{179}}{4}^{(80)}$ ,  $\frac{\sqrt{83}}{4}^{(120)}$ 
\\[4pt]  \hline \hline
\multirow{4}{*}{$\cN=0$, G$_2$} & $0$    & $\frac{3}{\sqrt{2}}^{(1)}$ , $\sqrt{\frac{3}{2}}^{(7)}$          \\[4pt]  
\cline{2-3} %\cline{7} 
& $ 1$ & $\sqrt{\frac{21}{2}}^{(1)}$ ,   $\sqrt{\frac{15}{2}}^{(14)}$ ,   $\frac{3}{\sqrt{2}}^{(14)}$ ,    $\sqrt{\frac{7}{2}}^{(27)}$ 
 \\[4pt]  
\cline{2-3} %\cline{7} 
& $ 2$ & $\sqrt{\frac{31}{2}}^{(7)}$ ,    $\sqrt{\frac{25}{2}}^{(14)}$ ,   $\sqrt{\frac{23}{2}}^{(54)}$ , $\sqrt{8}^{(64)}$, $\sqrt{\frac{13}{2}}^{(77)}$

\\[4pt]  \hline 
\end{tabular}

%}

}\normalsize

\caption{\footnotesize{The KK gravitino spectra up to KK level $n=2$
about the massive type IIA AdS$_4$ solutions that uplift on $S^6$ from vacua of $D=4$ $\cN=8$ ISO(7) dyonically-gauged supergravity with at least SU(3) symmetry.}\normalsize}
\label{tab:KKGravitiniISO7}
\end{table}

\end{landscape}

\bibliography{references}

\end{document}